\documentclass{IEEEtaes}

\usepackage[colorlinks,urlcolor=blue,linkcolor=blue,citecolor=blue]{hyperref}
\usepackage{color,array,amsthm}
\usepackage{graphicx}
\usepackage[nolist]{acronym}
\usepackage[T1]{fontenc}
\usepackage[cmex10]{amsmath}
\usepackage{amssymb}
\usepackage{amsthm}
\usepackage{amsfonts}
\usepackage{algorithmic}
\usepackage{algorithm}
\usepackage{cite}
\usepackage{textcomp}
\IEEEoverridecommandlockouts
\usepackage{gensymb}
\usepackage{times}
\usepackage[caption=false]{subfig}
\usepackage{paralist}
\usepackage{mathrsfs}
\usepackage{graphicx}
\usepackage{tabularx}
\usepackage{url}
\usepackage{mathbbol}
\usepackage{booktabs}
\usepackage{verbatim}

\usepackage{bm}            
\usepackage{xcolor}
\usepackage{multirow}
\usepackage{multicol}
\usepackage{cuted}
\usepackage{stfloats}

\expandafter\def\expandafter\UrlBreaks\expandafter{\UrlBreaks
    \do\a\do\b\do\c\do\d\do\e\do\f\do\g\do\h\do\i\do\j%
    \do\k\do\l\do\m\do\n\do\o\do\p\do\q\do\r\do\s\do\t%
    \do\u\do\v\do\w\do\x\do\y\do\z\do\A\do\B\do\C\do\D%
    \do\E\do\F\do\G\do\H\do\I\do\J\do\K\do\L\do\M\do\N%
    \do\O\do\P\do\Q\do\R\do\S\do\T\do\U\do\V\do\W\do\X%
    \do\Y\do\Z\do\/\do-}

\usepackage{tabularx, booktabs}

\begin{document}
\title{Design and Optimization of Aerial-Aided Multi-Access Edge Computing towards 6G} 
\author{von Mankowski, J\"org}
\affil{Technical University of Munich, Germany} 

\author{Durmaz, Emre}
\affil{Technical University of Munich, Germany}

\author{Papa, Arled}
\affil{Technical University of Munich, Germany} 

\author{Vijayaraghavan, Hansini}
\affil{Technical University of Munich, Germany} 

\author{Kellerer, Wolfgang}
\affil{Technical University of Munich, Germany}



\authoraddress{
J\"org von Mankowski is with the Technical University of Munich, 80333 Munich, Germany (joerg.von.mankowski@tum.de).
Emre Durmaz is with the Technical University of Munich, 80333 Munich, Germany (emre.durmaz@tum.de).
Arled Papa is with the Technical University of Munich, 80333 Munich, Germany (arled.papa@tum.de).
Hansini Vijayaraghavan is with the Technical University of Munich, 80333 Munich, Germany (hansini.vijayaraghavan@tum.de).
Wolfgang Kellerer is with the Technical University of Munich, 80333 Munich, Germany (wolfgang.kellerer@tum.de).}


\maketitle
\begin{abstract}
Ubiquity in network coverage is one of the main features of 5G and is expected to be extended to the computing domain in 6G. In order to provide this holistic approach of ubiquity in communication and computation, an integration of satellite, aerial and terrestrial networks is foreseen. In particular, the rising amount of applications such as In-Flight Entertainment and Connectivity Services (IFECS) and SDN-enabled satellites renders network management more challenging. Moreover, due to the stringent Quality of Service (QoS) requirements edge computing gains in importance for these applications. Here, network performance can be boosted by considering components of the aerial network, like aircrafts, as potential Multi-Access Edge Computing (MEC) nodes. 
Thus, we propose an Aerial-Aided Multi-Access Edge Computing (AA-MEC) architecture that provides a framework for optimal management of computing resources and internet-based services in the sky.
Furthermore, we formulate optimization problems to minimize the network latency for the two use cases of providing IFECS to other aircrafts in the sky and providing services for offloading AI/ML-tasks from satellites. 
Due to the dynamic nature of the satellite and aerial networks, we propose a re-configurable optimization. For the transforming network we continuously identify the optimal MEC node for each application and the optimal path to the destination MEC node. In summary, our results demonstrate that using AA-MEC improves network latency performance by 10.43\% compared to the traditional approach of using only terrestrial MEC nodes for latency-critical applications such as online gaming. Furthermore, while comparing our proposed dynamic approach with a static one, we record a benefit of at least 6.7\% decrease in flow latency for IFECS and 56.03\% decrease for computation offloading.
\end{abstract}


\section{Introduction}
\label{sec:Introduction}

\ac{6G} mobile networks aim to provide global coverage to ensure availability and seamless access \cite{bernardos2021european}. This not only enables communication in sparsely populated and isolated regions but also supports new applications like autonomous mobility, for example, autonomous ships and trains.
Terrestrial networks do not satisfy this requirement mainly due to the large amount of surface covered by water bodies and inaccessible locations on land. Moreover, given natural disasters and obstructed terrain, the terrestrial infrastructure cannot cope with the high availability demand. 
Therefore, satellite networks have been suggested to realize global coverage. However, satellites exhibit limitations due to their physical location. Space radiation requires hardware to be laid out redundantly with larger structural sizes compared to terrestrial hardware. This results in higher energy consumption and lower computing capacity. In addition, upgradability is constrained by the physical access possibilities exacerbating both problems over time. 
Furthermore, satellite networks exhibit the highest latency compared to other network types. In the best case, LEO satellites record a minimum round trip time of ca. 1~milliseconds due to a theoretical minimum operating altitude of 160~km. For geostationary satellites this increases to ca. 240~milliseconds, as they operate at 36000~km. These latency values fall far behind the \ac{6G}-envisioned latency requirements \cite{bernardos2021european}. 

Deploying \acp{ABS}, i.e. \acp{UAV}, drones, balloons, or aircrafts equipped with wireless transceivers providing \acp{ARAN} \cite{dao2021survey}, instead of satellites, overcomes such shortcomings. This is due to a flight altitude of maximum 10-15~km resulting in a latency of ca. 50~\micro s. At this altitude communication hardware is not affected by radiation anymore. Furthermore, \acp{ABS} offer better channel conditions compared to satellite networks, due to the propagation distance and a higher \ac{LoS} probability compared to terrestrial networks.
However, coverage area of a single \ac{ABS} is smaller than that of an satellite increasing the amount of handovers.

Combining the terrestrial, aerial and satellite network allows to harness the benefits and to compensate for the deficits of each network.
The overlapping coverage of these networks allows increasing the total available data rate, the user experienced data rate, and the device density. In addition, the redundancy introduced by overlapping coverage results in a more fail-safe and reliable network. This is especially important in cases of natural disasters, wars and other events that usually damage the local terrestrial infrastructure and lower the reliability of terrestrial networks. Moreover, this \ac{MLN} allows the usage of different layers for different applications. The satellite layer reduces handovers for globally moving nodes, whereas the terrestrial layer exhibits the highest data-rate for local nodes, resulting in a better total \ac{QoS} for every node in the network. 
Also, cost concerns that previously benefited the deployment of terrestrial networks over other network infrastructures have lost significance as satellite deployment costs have shrunk due to increased subsidies, technological advancements and more competition \cite{jones2018recent}. Arguments that terrestrial networks are easier to maintain render obsolete, as the physical distance of satellite networks lead to different maintenance costs. Examples for profitable, usable and large-scale satellite networks as of today are Kuiper, Starlink and OneWeb. In addition, previously mentioned new applications make combined network approaches more independent of market demands generated by population density.
In summary, an \ac{MLN} is able to increase the capacity, the user experienced data rate and energy efficiency, and supports a higher node density while decreasing the latency, which is in line with the goals of \ac{6G} \cite{bernardos2021european}. 

The considered \ac{MLN} also allows for ubiquitous and edge computing, by utilizing aircrafts as potential \ac{MEC}-nodes thereby placing the processing and storage capability throughout the network. This \ac{AA-MEC} facilitates the newly envisioned \ac{6G} applications as \ac{IoT}, Industry 4.0, Smart Cities and new AI/ML-based distributed applications\cite{bernardos2021european}. These applications require that processing tasks are completed with a maximum tolerable delay. Thus, tasks need to be offloaded taking calculation time and latency of the offloading process into account. Therefore, it is necessary to optimize the placement of the edge computing units and the routing of the task to the edge. 

\begin{figure}[t]
  \centering
  \includegraphics[width=0.9\columnwidth]{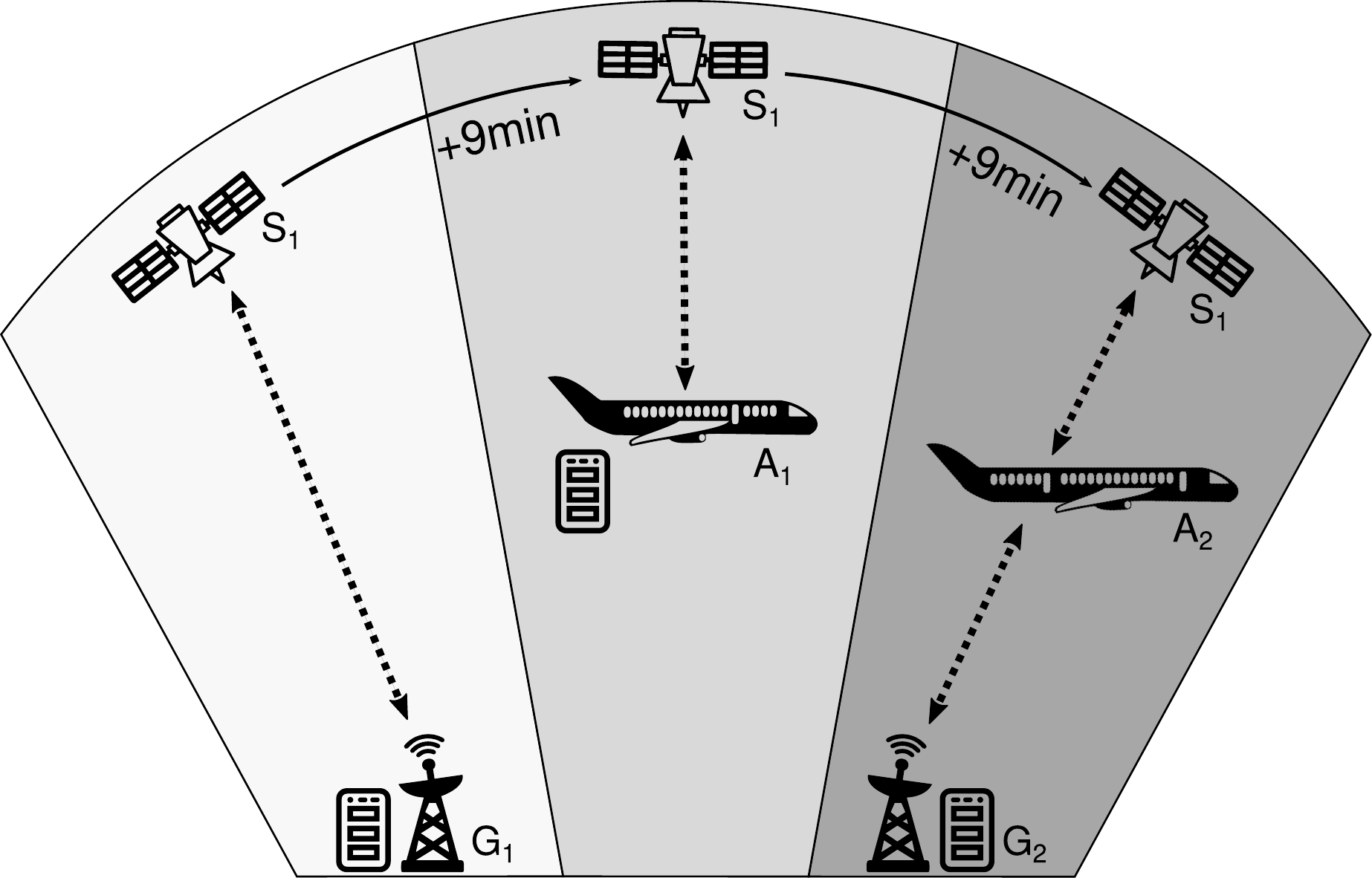}
  \caption{Dynamically changing \acl{MLN} topology over time}
  \label{fig:dynamic_links}
\end{figure}

In an \ac{MLN} most nodes are mobile, e.g. satellites and aircrafts. These nodes exhibit different speeds and trajectories. Fig.~\ref{fig:dynamic_links} visualizes this by showing how the links and \ac{MEC} destinations of a satellite change while orbiting the earth.
This results in continuously varying distances between nodes and dynamic changes in the network topology. This raises new challenges in routing, placement of computing tasks, and the provision of \ac{QoS}, especially latency guarantees. The dynamicity of the nodes influences the existence of a route and its life span. As trajectories and links change in real time, routes cannot be planned in advance. In this scenario, a static routing approach would result in increasing latency or even breaks in communication links. In computational task offloading applications, a static selection of \ac{MEC} destinations detracts from its purpose since the computing instances are not always on the edge to the changing network topology. The adaptability results in the need to reconfigure to always maintain optimality in path results in a more complex optimization problem. This results in the need for a new optimization problem description wherein the network must be optimized dynamically.

\subsection{Contribution of this study}
The requirements for 6G networks, particularly in terms of latency and the need for seamless integration of the satellite, aerial and terrestrial networks, solutions to reduce delay become imperative.
However, the design of such a system given the mobile nature of the envisioned three-dimensional architecture becomes challenging.
In this work, to cater for delay reduction, in line with 6G requirements, we propose an AA-MEC architecture, where already existing aircrafts can serve as computing entities for other components in the sky, for instance satellites or other aircrafts.

In this regard, we focus on two use cases. The first provides \acp{IFECS} to aircrafts during flights. The second enables processing for tasks offloaded from satellites, e.g. tasks related to SDN-enabled or AI-enabled satellite computations. We formulate and solve optimization problems with the goal of minimizing network latency for these two use cases while guaranteeing a maximum packet delay requirement for all the services. Moreover, as part of the optimization, our proposal defines the optimal edge computing destination (aerial or terrestrial) for each aircraft or satellite as well as the optimal path towards the destination. Due to the network dynamicity, we consider a re-configurable allocation of the processing resources to ensure that at each network change, the optimal solution is determined.

\subsection{Article Structure}
The remainder of this article is structured as follows. A detailed analysis of the related literature is given in Section~\ref{sec:related_work}.  Further,  the technologies used for the different network components are elaborated in Section~\ref{sec:background}. The system model is described in Section~\ref{sec:system_model}, whereas the optimization problem formulation is described in Section~\ref{sec:general_problem_formulation}. The main findings of this work are illustrated in Section~\ref{sec:performance_evaluation}. Finally, the implications of the \ac{AA-MEC} are discussed in Section~\ref{sec:conclusion}.

\section{Related work}
\label{sec:related_work}
The state of the art that is relevant for our proposed solution comes from three main areas: airborne internet, computational offloading and architectural concepts.
\subsection{Airborne Internet}
While many publications consider ground users as the source of service requests in \ac{MLN} \cite{chen2017caching} \cite{pacheco2021towards}, it is important to consider aircrafts as sources because airborne internet has been growing in demand in recent years. This is particularly a challenge due to the high speed of the nodes and the resulting rapid changes in the network topology and channel conditions.

The authors in \cite{Medina2010} consider an aerial mesh network that requires airborne internet services. The requested content is transmitted from a satellite gateway on the ground to the nearest aircraft and then routed to the other aircrafts requesting this content, through a multi-hop aerial network. The routing is optimized and this results in reduction of latency since the aircrafts can act as relays for the content from the ground station.
Similarly, the focus of \cite{Varasteh2019Mobility} is on providing airborne internet for a European Space-Air-Ground-Integrated Network. They also consider that the services are placed on ground, in data centers. The study compares different optimization algorithms focusing on routing, service placement, and service migration.

In contrast to the above-mentioned works, the authors in \cite{chen2021reinforcement} consider caching the content also on aircrafts and adding communication links between aircrafts. To solve the more complex routing problem they apply a reinforcement learning based algorithm which optimizes for energy efficiency. However, they do not take into account latency requirements for \ac{QoS} although aircraft passengers can request latency critical applications. Moreover, the dynamicity of the links leading to connection losses and increasing latency are also not considered. This new, more complex optimization is taken into account in our work by continuously optimizing the \ac{MEC} destinations for service placement while also providing latency guarantees. 

\subsection{Computation Offloading}
With the emergence of processing intensive applications like AI/ML, the importance of computational offloading increases. For this, various studies investigated offloading schemes in \ac{MEC}. 
Reference \cite{Baek2019} applies a reinforcement learning approach to minimize latency in Fog networks, a special case of \ac{MEC} networks. However, the latency constraints of Fog networks limit the placement of task offloading destinations. Furthermore, the dynamicity of the network is not considered and routes are not recalculated, which would be important in a dynamic \ac{MLN}.

An architecture that deploys \ac{MEC} nodes in a satellite and terrestrial network is considered in \cite{Wang2018}. They describe an offloading algorithm with the goal to optimize energy consumption and reduce latency by assigning computational tasks at minimal cost. For this, they consider special satellites as relay stations. However, they do not consider terrestrial or aerial \ac{MEC} nodes, which would add complexity to the problem.

The practicality of deploying \ac{MEC} nodes on satellites is limited due to increased latency and reduced calculation capabilities for hardware deployed in space. In \cite{Cheng2019}, an \ac{MLN} that consists of a \ac{LEO} constellation, \acp{UAV} and \ac{IoT} devices is studied. The network uses \acp{UAV} as flying \ac{MEC} servers and proposes a task scheduling mechanism for that architecture. The users have computational tasks to offload to the \acp{UAV} carrying \ac{MEC} servers and the computing resource allocation and task scheduling is optimized for this scenario. This optimization is solved by applying reinforcement learning. However, they do not consider connections between terrestrial gateways and aircrafts. In addition, aerial nodes are just considered to provide computational resources but not act as endpoints for content requests.

Reference \cite{Zhang2019} proposes an architecture called satellite \ac{MEC}, which is a \ac{MEC} service offered via satellite links. This service can be provided in different scenarios wherein the location of the \ac{MEC} server is either the satellite or the terrestrial station. Terrestrial offloading works similar to regular \ac{MEC} schemes, since servers are located in close proximity. On the other hand, satellite \ac{MEC} servers are deployed in LEO satellites in satellite-borne offloading. The limitation of this method is the resulting increased energy consumption of a satellite and the inability to deal with latency-critical applications. 
The authors of \cite{yu2021ec} propose a \ac{MLN} network where the satellite, aircraft and terrestrial layers contain \ac{MEC} servers and tasks request only  originate from the terrestrial layer. For this network architecture they propose a deep imitation learning-driven offloading and a caching algorithm to find the optimal task and cache placement. 
However they do not consider computational task requests from other layers.

In contrast to the above-mentioned works, firstly we extend the \ac{MEC} capability to aircrafts, thus bringing computational resources and content services closer to the origin of the requests in the air. Secondly, we also take the dynamicity of the network into account while selecting the destinations for offloading tasks by performing a re-configurable optimization.

\subsection{Architecture}
To enable latency-critical applications and provide a comprehensive network that supports \ac{6G} requirements and aerial use cases, the design of an optimal \ac{MLN} architecture is of great importance.
Inmarsat ORCHESTRA \cite{inmarsat_webpage} envisions a dynamic multi-layered network architecture consisting of satellite and terrestrial layers to support 5G. They aim to provide global network coverage for mobile nodes and enable new use cases like Smart Ships and connectivity for Urban Air Mobility. However, they only focus on connectivity solutions and ignore the potential of the network elements to act as computing entities. They also do not consider an aerial layer as part of this architecture. Moreover, even in such a 2-layer network providing connectivity solutions, the network can be optimized further for performance metrics such as latency. 

The authors in \cite{Chen2021} optimize the placement of the terrestrial gateways for SpaceX's Starlink as the reference constellation in the satellite layer. For this, the three metrics of latency, maximum load on the node, and load balancing among gateways are considered. The placement coordinates are obtained by applying a genetic algorithm. 
However, such a free optimization is not always feasible since the positioning of gateways is also influenced by political and economic considerations.
Networks with constraints on gateway positions and an additional aerial layer containing \ac{MEC} servers, as in our case, are therefore more difficult to optimize.

An aerial layer is incorporated in \cite{Zhen2021} resulting in an \ac{MLN} with an aerial layer consisting of short range \acp{UAV} and a terrestrial layer with users and base stations. The users have computational tasks that could be offloaded to the \ac{MEC} servers on the base stations or \acp{UAV}. However, with the increasing applications of AI/ML on satellites resulting in computationally intensive tasks, it becomes imperative to also consider a satellite layer in the design of \acp{MLN}. Besides, the authors perform an optimization to reduce the energy consumption in the network and calculate the optimal placement of tasks, trajectory of \acp{UAV} and offloading routes. Despite latency being a critical motivator for \ac{MEC} use cases, the authors do not provide a guarantee for the maximum latency for the offloaded tasks. They also do not consider any tasks arising within the aerial layer which could be served by other nodes in the aerial layer. Taking this factor into account would result in a highly dynamic network topology, thus adding complexity to the optimization problem. 

Therefore, in our \ac{MLN}, we include an aerial layer and a satellite layer that can act as sources for content and task requests, which is necessary given the rise in web services and computationally intensive applications in the sky. This results in a highly dynamic network which we also factor in while performing our re-configurable optimization. Since this dynamic \ac{MLN} is optimized for latency, we provide guarantees on the maximum delays experienced by the data packets, thus assuring \ac{QoS}.

\section{Background}
\label{sec:background}
As a basis for our system model that combines \ac{MEC} and \ac{MLN} we provide background information on Satellite Networks, \ac{MEC} and \ac{MLN}.
\subsection{Satellite Constellation Networks}
A satellite constellation consists of multiple satellites that orbit a planet to serve a common purpose. Iridium, Intelsat, Kuiper, OneWeb and StarLink are operational examples with the goal of providing a worldwide communication network coverage. These satellites may orbit at different trajectories and altitudes. One shortcoming of the mentioned satellite constellations is that they only provide a data rate of 25~Mbit/s as long as their user density does not exceed 0.1~user/m\textsuperscript{2} \cite{Osoro2021}. On the other hand, they are highly mobile and thus, more applicable than terrestrial networks in extreme situations. For instance, satellites have been successfully applied to enable operational networks after heavy damage to terrestrial infrastructures due to military conflicts and natural disasters.
Moreover, applying satellites may support terrestrial networks whose access is highly constrained due to environmental factors, e.g. undersea cables.
On top of that, technological advancements may provide for additional applications of satellites, e.g. intelligent transportation, remote area monitoring, disaster rescue, and large-scale high-speed mobile internet access \cite{Liu2018}. The constellation design not only influences the \ac{QoS}, measured by communication latency, handover frequency, throughput but also metrics such as manufacturing cost, orbital period, the number of satellites in the constellation and routing complexity.

We focus on \ac{LEO} satellites with an altitude range of 160 to 2,000~km, as they exhibit the lowest communication latency, the highest relative ground speed resulting in the highest flexibility. This is in line with the current trend to deploy LEO satellite constellations \cite{Saeed2021}, due to the recent technological advances and increasing importance of latency-sensible applications. However, the benefits of \ac{LEO} satellites come at the cost of a smaller coverage hence higher handover frequency. In addition, the high amount of satellites allow multiple flows which makes the traffic routing more flexible. This is particularly advantageous for the use as a backbone network. However, this flexibility leads to frequent changes in the network topology and therefore an increased routing complexity.

The specific network we base our results on is the Iridium-Next satellite constellation network \cite{IridiumConstellation} with 66 satellites operating at an altitude of 781~km. Satellites are grouped into multiple orbital planes resulting in \acp{ISL} spanning intra- and inter-plane communication. 

A large proportion of the relevant communication endpoints are located in the terrestrial network. Therefore, it is important to enable non-terrestrial networks to access these via terrestrial gateways. In general, satellites are served by one gateway at a time \cite{Chen2021}. As the placement of gateways directly influences the performance of the designed architecture \cite{Del2018}, gateways should be placed near heavily used terrestrial nodes, but also throughout the globe, to ensure global coverage, reliability and minimal latency. In case of the Iridium-Next Satellite network, gateways are not distributed evenly around the world, but can mostly be found in regions of heavy usage like North America and Eurasia. An overview of the gateways is provided in Table~\ref{tab:IridiumGateways}. 
\begin{table}[t]
\caption{Gateways of Iridium-Next network \cite{iridium_gateways}}
\label{tab:IridiumGateways}
\begin{center}
\begin{tabular}{p{0.3\columnwidth}p{0.15\columnwidth}p{0.15\columnwidth}p{0.15\columnwidth}p{0.15\columnwidth}}
\toprule
\textbf{Gateway Location} &\textbf{Country} &\textbf{Latitude (degrees)} &\textbf{Longitude (degrees)} \\
\midrule
Beijing  & China & 39.92 & 116.388 \\ 
Fairbanks  & USA & 64.838 & -147.716 \\ 
Iqaluit  & Canada & 63.733 & -68.500 \\ 
Ischewsk  & Russia & 56.850 & 53.204 \\ 
Longyearbyen  & Norway & 79 & 17.66 \\ 
Punta Arenas  & Chile & -53.315 & -71.580 \\ 
Rome  & Italy & 41.9 & 12.483 \\ 
Tempe  & USA & 33.415 & -111.909 \\ 
Wahiawa  & USA & 21.503 & -158.024 \\ 
Yellowknife & Canada & 62.450 & -114.350 \\ 
 \bottomrule
\end{tabular}
\end{center}
\end{table}
These satellite gateways can act as \ac{MEC} nodes for different processing tasks in the network.

\subsection{\acf{MEC}}
It can be expected that network nodes vary in processing and storage capabilities. This results in the possibility of computing tasks not being feasibly solved locally. To alleviate this problem, \ac{MEC} is introduced.
\ac{MEC} is a special case of edge computing proposed by the \ac{ETSI} \cite{etsi_arch}. This involves placing real-time or near-real-time processing servers at edge nodes within the \ac{RAN}. Computational tasks can be offloaded to \ac{MEC} servers, thereby enabling the movement of tasks between network nodes. This introduces a latency due to the propagation time and the execution time of the task. Hence, typically, \ac{MEC} nodes are located in proximity to end users which is the key factor to decrease serving latency.
The \ac{MEC} concept has become quite popular in recent years with applications like \ac{IoT}, data caching, video analytics, and autonomous driving \cite{ETSI_MEC}.
Therefore it is expected to play a key role in \ac{6G} communication systems.
However, binding \ac{MEC} servers to gateways, immobilizes them and therefore reduces their usage locations and applications, for example, in sparsely populated or remote areas or military, emergency relief, and disaster response. 
In cases when a computational task can be executed on multiple \ac{MEC} nodes within the network, the placement of it emerges as an optimization problem. 

\subsection{\acl{MLN}s}
In this work, the network under study is a \acl{MLN} comprising a terrestrial, an aerial and a satellite network. Base stations, satellite and \ac{DA2G} gateways are nodes found in the terrestrial layer. Aircrafts, low and high-altitude platforms and in general \acp{UAV} are found in the aerial layer and satellites are located in the satellite layer. Each layer provides certain resources and displays technological limitations. For instance, inter-satellite communication is often based on optical links which, on one hand, require a LOS channel, but on the other hand, exhibit high bandwidth compared to radio based satellite-gateway links. As another example, aircrafts may provide a high-speed access network with \ac{LEO} satellites being used as relay stations. The high mobility of the nodes in the aerial and satellite layers results in a dynamic network topology which allows for a more flexible network. An overview of these resources and limitations is given in Table~\ref{tab:LayerComparison}. The combination of the different types of networks allow each layer to compensate for the weaknesses of the other, resulting in the best use of the 3D space \cite{Wang2021ntn6G}. 
It is expected that this kind of network will be standardized in \ac{6G} \cite{Giordani2021}. Such \ac{6G} networks need data streams to be routed with minimal latency to support the various applications. However, the different communication media combined with the fast moving nodes in the non-terrestrial layers give rise to a new research challenge of how to optimally route data throughout the network. 

\begin{table}[t]
\caption{Layer comparison in an \ac{MLN}}
\label{tab:LayerComparison}
\begin{center}
\setlength\tabcolsep{4pt}
\begin{tabular}{lccc}
\toprule
  \textbf{Layer} &  \textbf{Nodes} &  \textbf{Advantages} &  \textbf{Disadvantages} \\
\midrule
Satellite & \begin{tabular}{ c } GEO, \\ MEO, \\ LEO \end{tabular} & \begin{tabular}{ c } large coverage, \\ infrastructure \\independent  \end{tabular}  & \begin{tabular}{ c }  LoS channel, \\ propagation delay  \end{tabular}\\
\midrule
  Aerial & \begin{tabular}{ c } HAP,\\ LAP,\\ UAV,\\ Aircraft \end{tabular}     &\begin{tabular}{ c }wide coverage,\\flexible de-\\ployment,\\low cost \end{tabular}  & \begin{tabular}{ c } unstable link,\\high mobility,\\less capacity  \end{tabular}\\ 
\midrule
Terrestrial &  \begin{tabular}{ c }  Cellular,\\ WiFi \end{tabular} & \begin{tabular}{ c } rich resources,\\ high throughput  \end{tabular}  & \begin{tabular}{ c } limited coverage,\\ infrastructure\\ dependent  \end{tabular}\\ \bottomrule
\end{tabular}
\end{center}
\end{table}

\section{System Model and Use Cases}
\label{sec:system_model}

The integration of \ac{MEC} nodes into the \ac{MLN} adds the \ac{6G} requirement of network flexibility to the \ac{MEC} concept. We envision that this combined architecture results in an \ac{AA-MEC}.  
In this \ac{AA-MEC} we address two research challenges, the first being the optimal placement of processing tasks across the multiple layers of a dynamically changing network and the second being the optimal routing of the processing tasks to the \ac{MEC} server.

\begin{figure}[t]
  \centering
  \includegraphics[width=0.9\columnwidth]{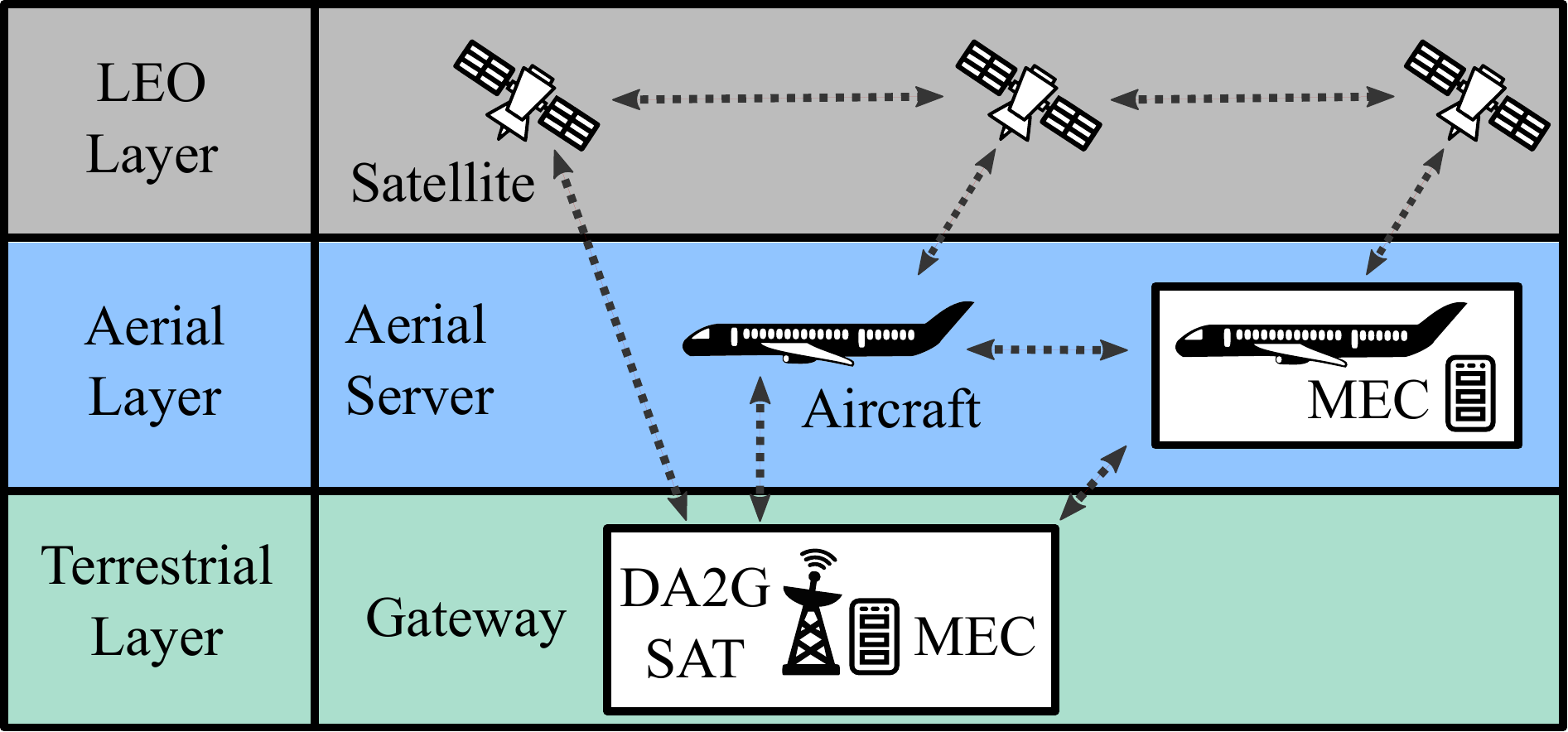}
  \caption{Designed multi-layer architecture.}
  \label{fig:DesignedMultiLayerArchitecture}
\end{figure}

The \ac{AA-MEC} architecture designed for this study is illustrated in Fig.~\ref{fig:DesignedMultiLayerArchitecture}.
The network under study consists of a \ac{LEO} Iridium-Next satellite, an aerial and a terrestrial layer. Nodes in the satellite layer possess processing capabilities, but this is limited and hence are not suitable recipients for processing tasks from other nodes. Satellites can communicate with terrestrial gateways and aircrafts found in the  aerial layer. Aircraft nodes can be further distinguished into passenger and aerial server aircrafts. While passenger aircrafts can function as routing nodes, server aircrafts also possess computing capability. Both aircraft types can communicate  with terrestrial gateways over air to ground links. All terrestrial nodes are able to communicate with satellite and aerial nodes. In all cases, \ac{MEC} servers are modeled as an integrated part of the network nodes of the \ac{RAN}.

A key requirement in a \ac{MEC} network is to minimize the user-perceived latency. This study aims to find the effects of the proposed multi-layer architecture on the user-perceived latency for the use cases of airborne internet and computation offloading.

\subsection{Use Cases}
\label{subsec:considered_use_cases}
\subsubsection{Airborne Internet}
\label{subsubsec:aec_for_aircrafts}
In recent years airborne internet connectivity is growing in demand. This allows critical information, such as weather, flight altitudes and landing conditions to be exchanged. Besides, it enables airplane passengers to receive web services such as messaging services, web-surfing, gaming, \ac{VoIP}, video and music streaming etc. The main obstacle to airborne connectivity is the lack of communication infrastructure, especially while flying over regions as deserts, the poles and large bodies of water. In such cases, the internet is traditionally provided over satellites resulting in high latency. This is even the case if a \ac{LEO} constellation is used. Aerial \ac{MEC} nodes and \ac{DA2G} can alleviate this problem especially in time-sensitive services. An overview of the service requirements can be found in Table~\ref{tab:ServiceProperties}.
\begin{table}[t]
\caption{Services provided through airborne internet \cite{etsi_spesification,savi2015impactservices,Garcia2018}}
\label{tab:ServiceProperties}
\begin{center}
\setlength\tabcolsep{1.3pt}
\begin{tabular}{lcccc}
\toprule
\textbf{Service Type} & \textbf{Bandwidth} &\textbf{Delay} &\textbf{Utilization Ratio}  &\textbf{Packet Size}  \\
\midrule
Web Service &  100 kbps & 500 ms & 0.14 &  933 B \\ 
Online Gaming &  50 kbps &  60 ms & 0.04&  24 B \\
VoIP & 64 kbps & 100 ms & 0.15&  829 B \\ 
Video Streaming &  1.5 Mbps &  300 ms &  0.67 &  1378 B \\ 
\bottomrule 
\end{tabular}
\end{center}
\end{table}
\subsubsection{Computation Offloading}
\label{subsubsec:aec_for_satellites}
Different computational tasks arise in satellites. In case the task exceeds the processing capability of the satellite, it can be offloaded to a \ac{MEC} server. This is defined as computational offloading. This comes at the cost of an additional propagation and transmission latency. However, deploying \ac{MEC} in aircrafts and satellite gateways brings the computing resources and other services closer to the satellite nodes resulting in reduced propagation and transmission time. As a result, deploying \ac{AA-MEC} would bring considerable gains to satellite networks. 
We assume that the satellite and \ac{MEC} server processors are ARM Cortex-A8 and ARM Cortex-A73 based, respectively. Table~\ref{tab:ProcessorSpecifications} gives an overview of their processing capabilities. 

\begin{table}[t]
\caption{Specifications of ARM processors~\cite{Jorgensen2020}}
\label{tab:ProcessorSpecifications}
\begin{center}
\setlength\tabcolsep{4pt}
\begin{tabular}{lccccc}
\toprule
\textbf{Processor Type} &\textbf{Frequency} &\textbf{IPC} &\textbf{Cores} &\textbf{MIPS} &\textbf{Tasks} \\
\midrule
ARM Cortex-A8 & 1 GHz & 2 & 1 & 2000 & 80 \\
ARM Cortex-A73 & 2.8 GHz & 6.35 & 4 & 71120 & 2844 \\
\bottomrule 
\end{tabular}
\end{center}
\end{table}

\section{Problem Formulation}
\label{sec:general_problem_formulation}
The two use cases result in two optimization problems with the goal of minimizing latency in an \ac{AA-MEC} where gateways and aircrafts act as \ac{MEC} nodes for \ref{subsec:aec_for_aircrafts_problem_formulation}) passenger aircrafts' \ac{IFECS} and \ref{subsec:aec_for_satellites_problem_formulation}) for satellites. All the trajectory simulations of satellites and aircrafts in this work were executed with the help of the \ac{STK} software \cite{agi_stk}.

The envisioned network topology is defined as a graph $\mathcal{G}$ = $(\mathcal{V}, \mathcal{E})$, where $\mathcal{V}$ are the network nodes (i.e., satellites, gateways and aircrafts). Further, the set $\mathcal{E}$ denotes the network edges of the communication links among all entities. The capacity of the links are given in Table~\ref{tab:link_capacity}.
\begin{table}[t]
\caption{Capacity of different link types in the network}
\label{tab:link_capacity}
\begin{center}
\begin{tabular}{lc}
\toprule
\textbf{Link Type} &\textbf{Capacity}   \\
\midrule
Satellite - Satellite & 125~Mbps \\ 
Satellite - Airplane  & 112~Mbps \\ 
Satellite - Gateway  & 500~Mbps \\ 
Airplane - Airplane &  45~Mbps \\ 
Airplane - Gateway (DA2G) &  75~Mbps \\ 
\bottomrule
\end{tabular}
\end{center}
\end{table} 
In the Iridium-Next constellation created with \ac{STK}, each satellite has two stable \acp{ISL} within the same orbit and two stable \acp{ISL} to the neighboring orbits. \acp{ISL} are part of the edges $\mathcal{E}$. However, due to the satellite and aircraft mobility, connections change over time. This is reflected in our work by generating a snapshot $r \in \mathcal{R}$ of the network every $5$ minutes. Within a snapshot the network is assumed static~\cite{papa2020design}.

\subsection{Aircrafts as Edge Computing Entities}
\label{subsec:aec_for_aircrafts_problem_formulation}
In the airborne internet use case we consider services provided to passengers within aircrafts that fall into the category of \ac{IFECS}. All notations used in this use case can be found in Table~\ref{tab:VariableTableAirborneInternet}. In an aircraft $a \in \mathcal{V}_\mathcal{A}^r$ with $N_a$ passengers, only $\rho_a = 0.2$~\cite{Varasteh2019Mobility} utilize \ac{IFECS}. Depending on the aircraft, $N_a$ ranges from 132 up to 853 passengers with the mode at 180. We distinguish four different services $m \in \mathcal{M}$, where $U_{m}$ denotes the utilization ratio of a service and  $B_{m}$ the corresponding bandwidth requirement per passenger. An overview of the different services and their requirements is given in Table~\ref{tab:ServiceProperties}.
\begin{table}[t]
	\caption{Variables for airborne internet services optimization}
	\label{tab:VariableTableAirborneInternet}
	\begin{center}
	\begin{tabular}{p{0.15\columnwidth}p{0.75\columnwidth}}
		\toprule
		\textbf{Variable} & \textbf{Description}\\
		\midrule
		$\mathcal{R}$ & Set of network snapshots\\
		$\mathcal{M}$ & Set of services in aircrafts\\
		$\mathcal{V}_\mathcal{S}^r$ & Set of satellites at snapshot $r \in \mathcal{R}$\\
		$\mathcal{V}_\mathcal{A}^r$ & Set of aircrafts at snapshot $r \in \mathcal{R}$\\
		$\mathcal{V}_\mathcal{G}^r$ & Set of gateways at snapshot $r \in \mathcal{R}$\\
		$\mathcal{V}_r$ & Ordered set of nodes $\mathcal{V}_r = \mathcal{V}_\mathcal{S}^r \cup \mathcal{V}_\mathcal{A}^r \cup \mathcal{V}_\mathcal{G}^r$ at snapshot $r \in \mathcal{R}$\\
		$\mathcal{E}_r$ & Set of links at snapshot $r \in \mathcal{R}$, where $\mathcal{E}_r = \mathcal{V}_r \times \mathcal{V}_r$\\
		$\mathcal{F}$ & Set of flows, where each flow $f \in \mathcal{F}$ corresponds to a service $m \in \mathcal{M}$ and aircraft $a \in \mathcal{V}_\mathcal{A}^r$\\
		$B_{i,j,r}$ & Bandwidth of link $(i,j) \in \mathcal{E}_r$\\
		$D_{f_{a}^{m}}$ & Data of flow $f \in \mathcal{F}$\\
		$P_{f_{a}^{m}}$ & Size of a service packet within flow $f \in \mathcal{F}$	\\
		$\tau_{f_{a}^{m}}$ & Delay constraint of a packet within flow $f \in \mathcal{F}$\\
		$N_a$ & Total number of passengers in aircraft $a \in \mathcal{V}_\mathcal{A}^r$\\
		$\rho_a$ & Ratio of passengers that utilize services in an aircraft $a \in \mathcal{V}_\mathcal{A}^r$\\
		$U_m$ & Ratio of passengers that utilize service $m \in \mathcal{M}$\\
		$B_m$ & Bandwidth requirement of service $m \in \mathcal{M}$\\
		$u_{f_{a}^{m}, i, j}$ & 
		Binary variable indicating if link $(i,j) \in \mathcal{E}_r$ is used for flow $f$\\
		$x_{f_{a}^{m}}^{d}$ &
		Binary variable indicating if flow $f$ matches the destination $d \in \mathcal{V}_r$ \\
		$q_{s, g}$ &
		Binary variable indicating if satellite $s \in \mathcal{V}_\mathcal{S}^r$ has a connection with gateway $g \in \mathcal{V}_\mathcal{G}^r$ \\
		\bottomrule
	\end{tabular}
	\end{center}
\end{table}

In our scenario, each service within each aircraft is considered as network flow denoted as $f_{a}^{m}$. We can then write the total flow demand of a service as
\begin{equation}
D_{f_{a}^{m}} = B_{m} \cdot U_{m} \cdot N_a \cdot \rho_a.
\end{equation}
The total latency of a flow is the aggregation of propagation and flow transmission latencies. The flow transmission latency is defined as the time taken to transmit the whole flow demand ($D'_{f_{a}^{m}}$) in one second from the source to the destination and is expressed as
\begin{equation}
    {L'}_{f_{a}^ {m},i,j, r} = \frac{D'_{f_{a}^{m}} }{B_{i, j, r}},
\end{equation}
where $B_{i,j, r}$ is the bandwidth of the link that connects node $i$ with node $j$ at a given snapshot $r$. 
The transmission latency for a single packet within a flow $f$ can be written as
\begin{equation}
    {L'}^p_{f_{a}^{m}, i,j, r} = \frac{P_{f_{a}^{m}}}{B_{i,j, r}},
\end{equation}
where $P_{f_{a}^{m}}$ is the size of a packet. Note that we assume that every packet of a flow $f \in \mathcal{F}$ has the same size.
The propagation latency is defined as
\begin{equation}
    L^c_{i, j, r} = \frac{d_{i,j,r}}{c},
\end{equation}
where $d_{i,j,r}$ is the physical distance from node $i$ to node $j$ at snapshot $r$ and $c$ is speed of light. Hence, the total latency for a flow and a packet are provided as
\begin{align}
L_{f_{a}^{m}, i, j, r} &= {L'}_{f_{a}^{m}, i, j, r} + 2L^c_{i, j, r} \label{eq:flow_latency},\\
L^p_{f_{a}^{m}, i, j, r} &= {L'}^{p}_{f_{a}^{m}, i, j, r} + 2L^c_{i, j, r} \label{eq:packet_latency}.
\end{align}

Since our goal is to provide \ac{QoS} for \acp{IFECS}, we aim to minimize the total flow latency for all flows $f \in \mathcal{F}$ in the network, in all snapshots $r \in \mathcal{R}$, while guaranteeing a maximum delay for each packet within a flow $f$. To achieve this, the optimization problem focuses on two main aspects. The first concerns obtaining the optimal destination for each aircraft $a \in \mathcal{V}_\mathcal{A}^r$. This destination node can be either an aircraft or gateway denoted by $d \in  \mathcal{V}_\mathcal{A}^r \cup \mathcal{V}_\mathcal{G}^r$. The second aspect is to identify the shortest path to this destination. The overall formulation is given by
\begin{align} 
    \min_{u_{f_{a}^{m}, i, j}, x_{f_{a}^{m}}^{d}}\:\sum_{r \in \mathcal{R}}^{}\sum_{(i,j) \in \mathcal{E}_r}^{}\sum_{f \in \mathcal{F}}^{}&L_{f_{a}^{m}, i,j,r} \cdot u_{f_{a}^{m}, i, j} \label{eq:minimizationAirp},\\
    \mbox{s.t.} \sum_{d \in \mathcal{V}_\mathcal{A}^r \cup \mathcal{V}_\mathcal{G}^r} x_{f_{a}^{m}}^{d} = 1,\, \forall f &\in \mathcal{F} \label{eq:constSingleDestAirp},
\end{align}
\begin{align}
     \begin{split}
     \sum_{(i,k) \in \mathcal{E}_r}^{} u_{f_{a}^{m}, i, k} - \sum_{(k,j) \in \mathcal{E}_r}^{} u_{f_{a}^{m}, k, j} =\\
      \begin{cases}
    -1& \text{if } k = src_{f_{a}^{m}},\\
    x_{f_{a}^{m}}^{d} &\text{if } k \neq src_{f_{a}^{m}},\\ 
    \end{cases} \label{eq:constFlowConservationAirp}\\
    \end{split}
\end{align}    
\begin{align}
     \sum_{(i,j) \in \mathcal{E}_r}^{} 
     L^p_{f_{a}^{m}, i, j} \cdot u_{f_{a}^{m}, i, j} \leq &\tau_{f_{a}^{m}},\, \forall f \in \mathcal{F} \label{eq:constDelayAirp},\\
    \sum_{f \in \mathcal{F}}
    D_{f_{a}^{m}} \cdot u_{f_{a}^{m}, i, j} \leq B_{i, j, r}, &\forall (i,j) \in \mathcal{E}_r \label{eq:constBandwidthAirp} ,\\
    \sum_{i \in \mathcal{V}_\mathcal{S}^r}
    q_{i, j} \leq 1,\, &\forall j \in \mathcal{V}_\mathcal{G}^r \label{eq:GWSatSelectionConstAirp},\\
    \sum_{f \in \mathcal{F}} u_{f_{a}^{m}, i, j} = q_{i, j} \sum_{f \in \mathcal{F}} x_{f_{a}^{m}}^{j},& \forall (i,j) \in \mathcal{V}_\mathcal{S}^r \times \mathcal{V}_\mathcal{G}^r \label{eq:GWSatRestrictionConstAirp}.
\end{align}
Equation~\eqref{eq:minimizationAirp} expresses the minimization problem formulation given the constraints ~\eqref{eq:constSingleDestAirp} -~\eqref{eq:GWSatRestrictionConstAirp}. Here, the binary variable $u_{f_{a}^{m}, i, j}$ is equal to 1 if link $(i, j)$ is used for a flow $f$ of service $m$ and aircraft $a$ and 0 otherwise. Equation~\eqref{eq:constSingleDestAirp} ensures that each flows assigned to one server or destination, where $x_{f_{a, m}}^{d}$ is a binary variable that takes the value 1 if a flow $f \in \mathcal{F}$ is matched to its destination, else it remains 0.  Equation~\eqref{eq:constFlowConservationAirp} describes the flow conservation constraint for a generic node $k \in \mathcal{V}_r$ where the number of flows arriving to and leaving from a node must be equal unless $k$ is a source or destination node of flow $f \in \mathcal{F}$. Equation~\eqref{eq:constDelayAirp} guarantees that the packet delay of each flow remains below the maximum allowed value $\tau_{f_{a}^{m}}$. Equation~\eqref{eq:constBandwidthAirp} maintains the link bandwidth constraint. Further, \eqref{eq:GWSatSelectionConstAirp} allows at most one single satellite to be connected with a gateway $g \in \mathcal{V}_\mathcal{G}^r$, where $q_{i, j}$ is a binary variable that takes value of 1 if a satellite is selected to have a connection with a gateway and 0 otherwise. Finally, \eqref{eq:GWSatRestrictionConstAirp} ensures that flows from a satellite go to its connected gateway.  

\subsection{Edge Computing for Satellites}
\label{subsec:aec_for_satellites_problem_formulation}

In the case of satellite task offloading, we assume that each satellite $s \in \mathcal{V}_\mathcal{S}^r$ needs to process a specific amount of computational tasks. If the computational tasks surpass the processing capabilities of the satellite, the tasks will be offloaded to a \ac{MEC} node on a gateway $g \in \mathcal{V}_\mathcal{G}^r$ or on an aircraft $a \in \mathcal{V}_\mathcal{A}^r$. An overview of the additional notations used in this use case can be found in Table~\ref{tab:VariableTableOffloading}. The amount of tasks $J_s$ a satellite needs to process during a snapshot $r$ is modeled as a random variable following a Poisson distribution
\begin{equation}
    Pr(J_{s} = k)= \frac{\lambda_s^{k}e^{-\lambda_s}}{k!},
\end{equation}
where, $k$ is the number of arriving tasks and $\lambda_s$ is the average task rate for satellite $s$.
The processing capacity of each satellite $s$ is defined as the number of instructions its processor is able to handle in a second. The unit is given in \ac{MIPS} and calculated by
\begin{equation}
    C = Freq \cdot IPC \cdot n_{cores}.
\end{equation}
Here, $Freq$, $IPC$ and $n_{cores}$ correspond to the CPU frequency in cycles per second, instructions per cycle per core, and number of CPU cores, respectively. $C_{s}$ and $C^{MEC}$ are the processing capacity of a satellite $s$ and a \ac{MEC} server. 
A computational task consists of $I$ = $25\cdot 10^6$ instructions, processes $D_{T} = 0.2$ MB of data and should be completed in no longer than the delay requirement of $\tau_s = 1000$\,ms \cite{Jorgensen2020, Sthapit2021}.

The number of tasks offloaded by a satellite is calculated by
\begin{equation}
    O_{s} = J_{s} - C_{s}.
\end{equation}
The lower bound of the bandwidth needed to transmit the offloaded tasks of the satellite is given by
\begin{equation}
    O^B_{s} = \frac{D_{T}\cdot O_{s}}{\tau_s}.
\end{equation}
The satellite flow transmission latency is given as
\begin{equation}
    L_{s,i,j} = {L'}_{s, i, j} + 2L^c_{i, j},
\end{equation}
where ${L'}_{s, i, j}$,  denotes the task transmission latency over a link $(i,j) \in \mathcal{V}_{r} \times \mathcal{V}_{r} $ and $L^c_{i, j}$ the corresponding propagation latency.
The time a \ac{MEC} server requires to process the offloaded task is given by the computation latency as
\begin{equation}
    L_s^{MEC} = \frac{I\cdot O_{s}}{C^{MEC}} \label{eq:computation_latency}.
\end{equation}

The \ac{QoS} for the satellite computation offloading scenario is defined by the service latency which is the elapsed time from the request to the completion of the task~\cite{Rodrigues2017}. This is the aggregation of the round trip propagation latencies, transmission latency and computation latency introduced by the \ac{MEC} server.
\begin{table}[t]
	\centering
	    \caption{Variables for satellite offloading optimization}
	\label{tab:VariableTableOffloading}
	\begin{tabular}{p{0.15\columnwidth}p{0.75\columnwidth}}
		\toprule
		\textbf{Variable} & \textbf{Description}\\
		\midrule
		 $\lambda_{s}$ & Task arrival rate for satellite $s$\\
		 $IPC$ & Number of instructions per cycle per core \\
		 $Freq$ & CPU clock rate in cycles per second \\
		 $n_{cores}$ & Number of cores \\
		 $D_{T}$ & Size of an offloading task \\
		 $J_{s}$ & Number of incoming tasks for satellite $s$\\
         $O_{s}$ & Number of tasks to be offloaded from satellite $s$\\
         $O^B_{s}$ & Bandwidth required to offload a task from satellite $s$\\
         $\tau_s$ & Maximum time to complete the task for satellite $s$\\
         $C^{MEC}$ & Capacity of \ac{MEC} server in number of tasks \\
         $C_{s}$ & Capacity of satellite $s$ in number of tasks\\
         $u_{s, i, j}$ & 
          Binary variable indicating if a link $(i,j)$ is used for the task of satellite $s$\\
         $x_{s}^d$ &
         Binary variable indicating if the task of satellite $s$ is matched to destination $d \in \mathcal{V}_r$\\
      \bottomrule
	\end{tabular}
\end{table}

To enable satellite task offloading, we aim to minimize the task completion time over all snapshots $r \in \mathcal{R}$ for every satellite $s \in \mathcal{V}_\mathcal{S}^r$. The optimization problem identifies the optimal \ac{MEC} destination node for every satellite $s \in \mathcal{V}_\mathcal{S}^r$ and the shortest path to that node. The problem is formulated as
\begin{align}
    \min_{u_{s, i, j}, x_{s}^{d}}\:\sum_{r \in \mathcal{R}}^{}\sum_{(i,j) \in \mathcal{E}_r}^{} 
    &\sum_{s \in \mathcal{V}_\mathcal{S}^r}^{}
    L_{s, i, j,r} \cdot u_{s, i, j}  \label{eq:minimizationSat},\\
    \mbox{s.t.} \sum_{d \in \mathcal{V}_r}^{}
    x_{s}^{d} &= 1,\, \forall s \in \mathcal{V}_\mathcal{S}^r \label{eq:constSingleDestSat} ,\\
     \sum_{(i,k) \in \mathcal{E}_r}^{} u_{s, i, k} -
     \sum_{(k,j) \in \mathcal{E}_r}^{} u_{s, k, j} 
     &= \begin{cases}
    -1& \text{if } k = src_{s},\\
    x_{f_{a}^{m}}^{d} &\text{if } k \neq src_{s},\\ 
\end{cases} \label{eq:constFlowConservationSat}\\
     \sum_{(i,j) \in \mathcal{E}_r}^{} 
     L_{s, i, j} \cdot u_{s, i, j} + L_s^{MEC} &\leq \tau_s,\, \forall s \in \mathcal{V}_\mathcal{S}^r \label{eq:constNoBufferSat},\\
    \sum_{s \in \mathcal{V}_\mathcal{S}^r}^{}
    O^B_{s} \cdot u^{s}_{i, j} \leq B_{i,j} , \, &\forall (i,j) \in \mathcal{E}_r \label{eq:constBandwidthSat},\\
    \sum_{i \in \mathcal{V}_\mathcal{S}^r}
    q_{i, j} \leq 1,\, \forall j \in &\mathcal{V}_\mathcal{G}^r \label{eq:GWSatSelectionConstSat},\\
    \sum_{s \in \mathcal{V}_\mathcal{S}^r}^{}
    u_{s, i, j} = q_{i, j} \sum_{s \in \mathcal{S}}^{} x_{s}^{j}, \, &\forall (i,j) \in \mathcal{V}^r_{S} \times \mathcal{V}_\mathcal{G}^r. \label{eq:GWSatRestrictionConstSat} 
\end{align}
Equation~\eqref{eq:minimizationSat} gives the objective function to minimize task completion time, where $u_{s, i,j}$ is a binary variable that is equal to 1 if link $(i, j)$ is used by satellite $s$. Equation~\eqref{eq:constSingleDestSat} ensures that each satellite $s$ is assigned to one destination, where $x_{s}^{d}$ is a binary variable that takes value of 1 if a satellite $s$ is matched to a \ac{MEC} node. Equation~\eqref{eq:constFlowConservationSat} represents the flow conservation constraint. Equation~\eqref{eq:constNoBufferSat} guarantees that offloaded tasks are completed within $\tau_s$. Further, \eqref{eq:constBandwidthSat} guarantees that the sum of bandwidths for all tasks passing through a link $(i,j)$ does not exceed its limit. Finally, \eqref{eq:GWSatSelectionConstSat} ensures that a gateway is connected to at most one satellite and ~\eqref{eq:GWSatRestrictionConstSat} ensures that flows from a satellite go to its connected gateway. 

\section{Performance Evaluation}
\label{sec:performance_evaluation}
Extensive evaluations of our \ac{AA-MEC} model and optimization problem are based on the performance metric of flow latency (see \eqref{eq:flow_latency}). For this we consider different \ac{MEC} server deployment ratios throughout the aerial network. Specifically we examine network configurations with no aerial nodes containing a \ac{MEC} server, 20\% containing one or 40\%. Alternatively, all satellite gateways deploy a \ac{MEC} server in all configurations. To obtain the coordinates and trajectories of the nodes in the network we run the STK simulator taking a snapshot of the network every 5 minutes on the 18th of August 2021 from 12:00 to 16:00 UTC. This data is then passed to our optimizer, which solves the problem using Gurobi \cite{gurobi}. Results in the airborne internet use case are based on at least 147000 flows per MEC deployment ratio and in the satellite case on ca. 30000 flows per satellite task arrival rate.

\subsection{Static vs. Dynamic Approaches}
\label{results_static_dynamic}

We start our analysis by quantifying the difference between a static and a dynamic approach for both the airborne internet and satellite offloading use case. While in the dynamic setup, we find the optimal \ac{MEC} destination for every network snapshot and the optimal path in the network to reach that destination, in a static scenario the \ac{MEC} destination is optimally chosen only for the first snapshot and then remains static. Nonetheless, the optimal path to the destination is recalculated at every snapshot. This is necessary in case the path is lost due to the mobility of the nodes and in order to provide a fair comparison among the two approaches.

\begin{figure}[t]
\centering
    \subfloat[Comparison of different services for aircrafts]{%
    \label{fig:static_vs_dynamic_aircrafts_mec_2}
    \includegraphics[width=0.45\textwidth]{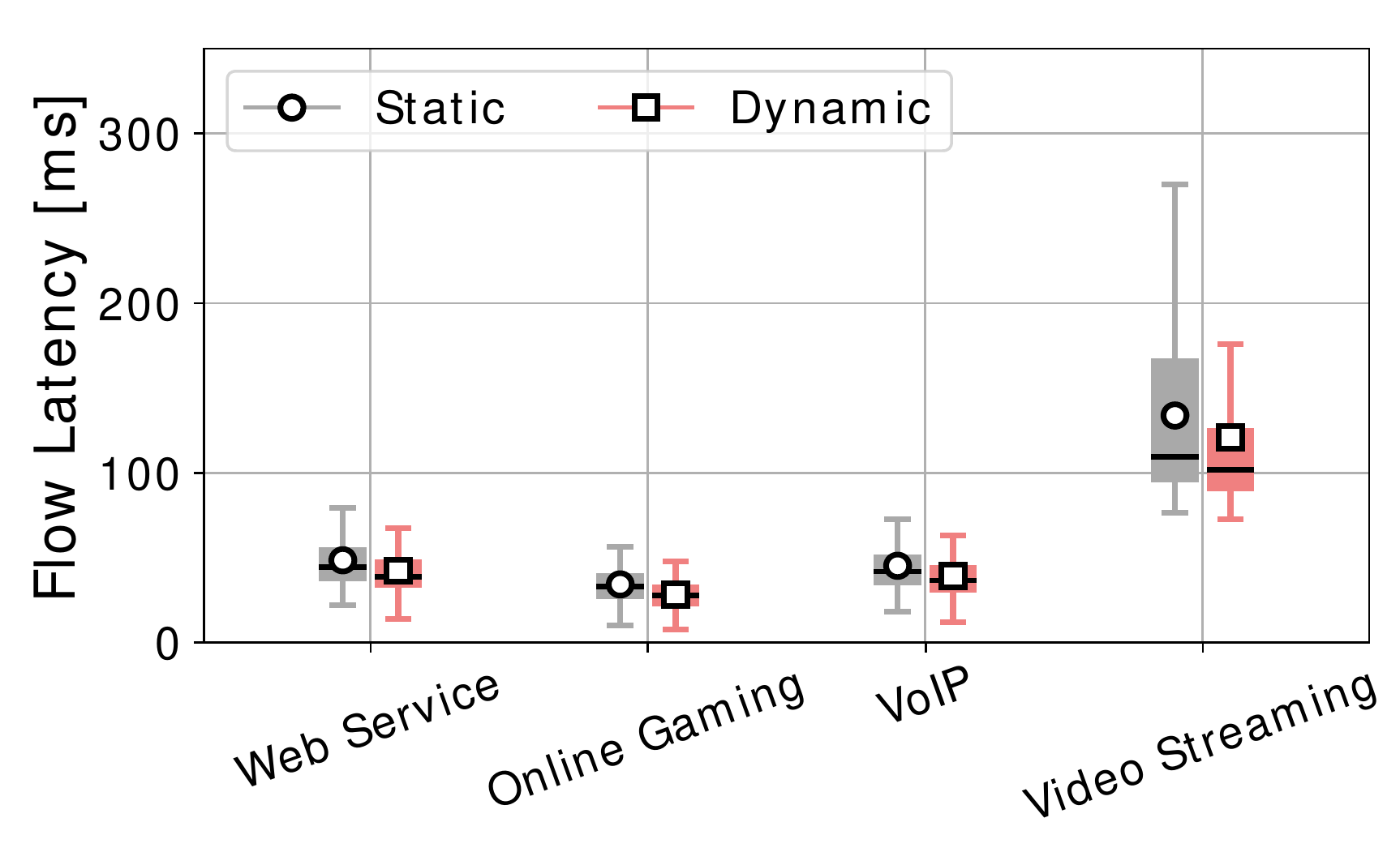}}
    \hfill
    \subfloat[Comparison of different task arrival rates for satellites]{%
    \label{fig:static_vs_dynamic_satellites_mec_2}
    \includegraphics[width=0.45\textwidth]{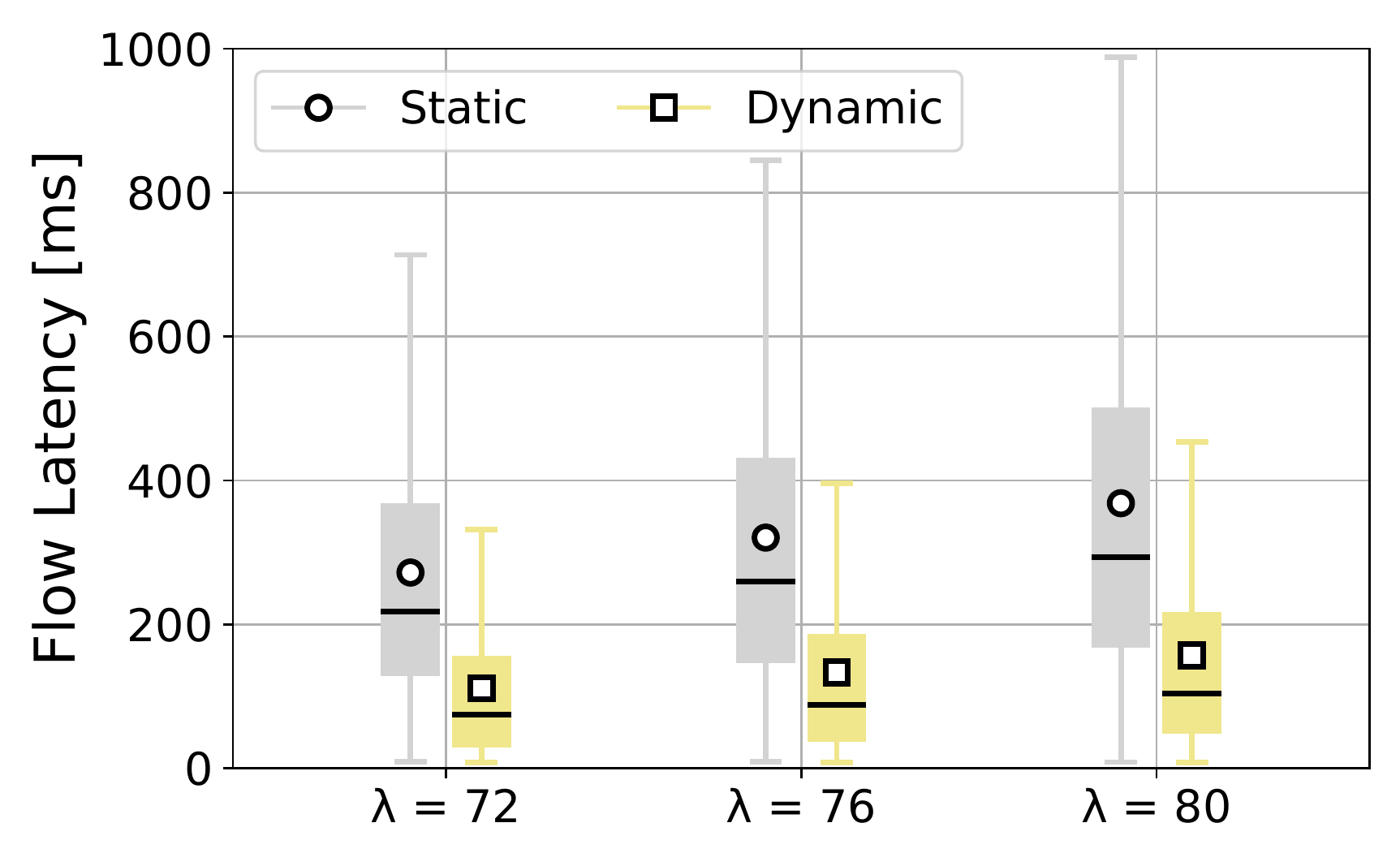}}
    \caption{Static vs. dynamic approaches for $\ac{MEC}_\text{20\%}$}
	\label{fig:static_vs_dynamic_approaches}
\end{figure}

Fig.~\ref{fig:static_vs_dynamic_aircrafts_mec_2} demonstrates this difference for four applications for the case of airborne internet, considering that 20\% of the nodes deploy a \ac{MEC} server. As can be seen from the figure, the dynamic approach outperforms the static for all applications in terms of flow latency. In web service, online gaming, VoIP and video streaming a 12.82\%, 16.3\%, 12.56\% and 6.7\% improvement can be seen, respectively. This is especially significant in gaming services where latency is a deciding factor for the outcome of the game and in VoIP applications where latency highly influences the \ac{QoE}.
Fig.~\ref{fig:static_vs_dynamic_satellites_mec_2} demonstrates the same trend for different task arrival rates $\lambda$ for the case of satellite task offloading. While the flow latency decreases by 59.44\% when $\lambda = 72$, the decreases become 58.71\% and  56.03\% for $\lambda = 76$ and $\lambda = 72$, respectively. Satellites benefit more from the dynamic approach as their higher relative ground speed results in larger increase in the distance to their \ac{MEC} node over time. The continuous optimization of the \ac{MEC} destinations mitigates this problem.
These results, substantiate the need to update selection of the \ac{MEC} destinations regularly.

\subsection{Airborne Internet} 
\label{results_airborne_internet}
Looking further into the implications of the optimization in the use case of the airborne internet we demonstrate the benefits of the \ac{AA-MEC} for aircrafts.
Table~\ref{tab:mec_utilization} shows that the utilization of satellite gateway \acp{MEC} decreases with an increase in aerial \ac{MEC} deployment ratio. This shows the benefit of the \ac{AA-MEC} which places \ac{MEC} nodes closer to the sources of services. For example, introducing 20\% of aerial \ac{MEC} nodes reduces the number of application flows ending in satellite gateway \acp{MEC} by 32\%. However, this effect begins to saturate for a higher number of aerial \ac{MEC} nodes, since the additional utilization drop is just 13\%. 
\begin{table}[t]
\caption{\ac{MEC} utilization for different aerial \ac{MEC} deployment ratios }
\label{tab:mec_utilization}
\begin{center}
\setlength\tabcolsep{4pt}
\begin{tabular}{lccc}
\toprule
\textbf{\ac{MEC} Type} &\textbf{$\ac{MEC}_\text{NO}$} &\textbf{$\ac{MEC}_\text{20\%}$} &\textbf{$\ac{MEC}_\text{40\%}$}  \\
\midrule
Gateway & 100\% & 68\% & 55\% \\
Aircraft &  0\% & 32\% & 45\% \\
\bottomrule 
\end{tabular}
\end{center}
\end{table}

\begin{figure}[th!]
\centering
    \subfloat[Web Service]{%
    \label{fig:individual_aircrafts_Packet_latency_Web_Service_service}
    \includegraphics[width=0.45\textwidth]{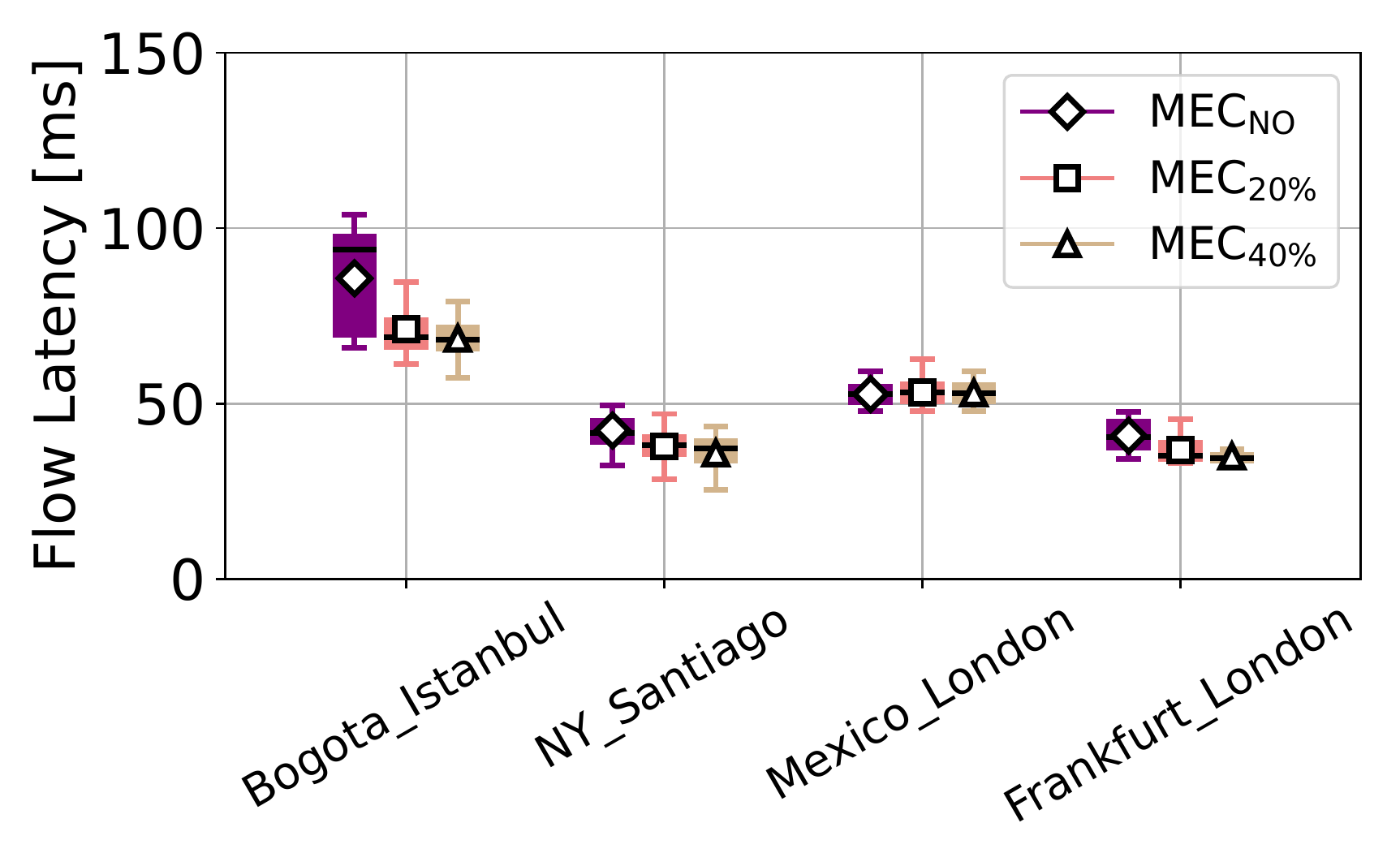}}\\
    \subfloat[Online Gaming]{%
    \label{fig:individual_aircrafts_Packet_latency_Online_Gaming_service}
    \includegraphics[width=0.45\textwidth]{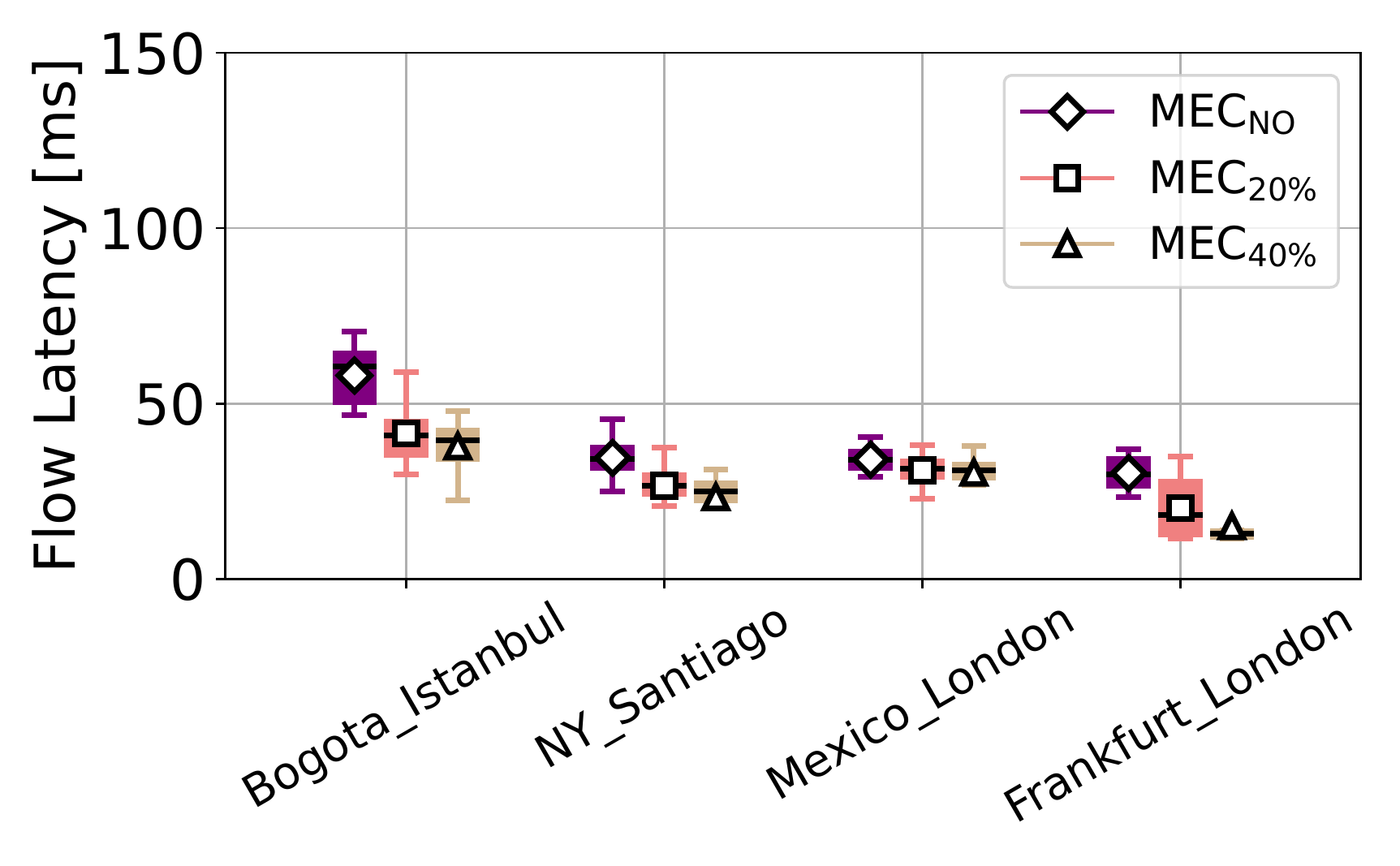}}\\
    \subfloat[VoIP]{%
    \label{fig:individual_aircrafts_Packet_latency_VoIP_service}
    \includegraphics[width=0.45\textwidth]{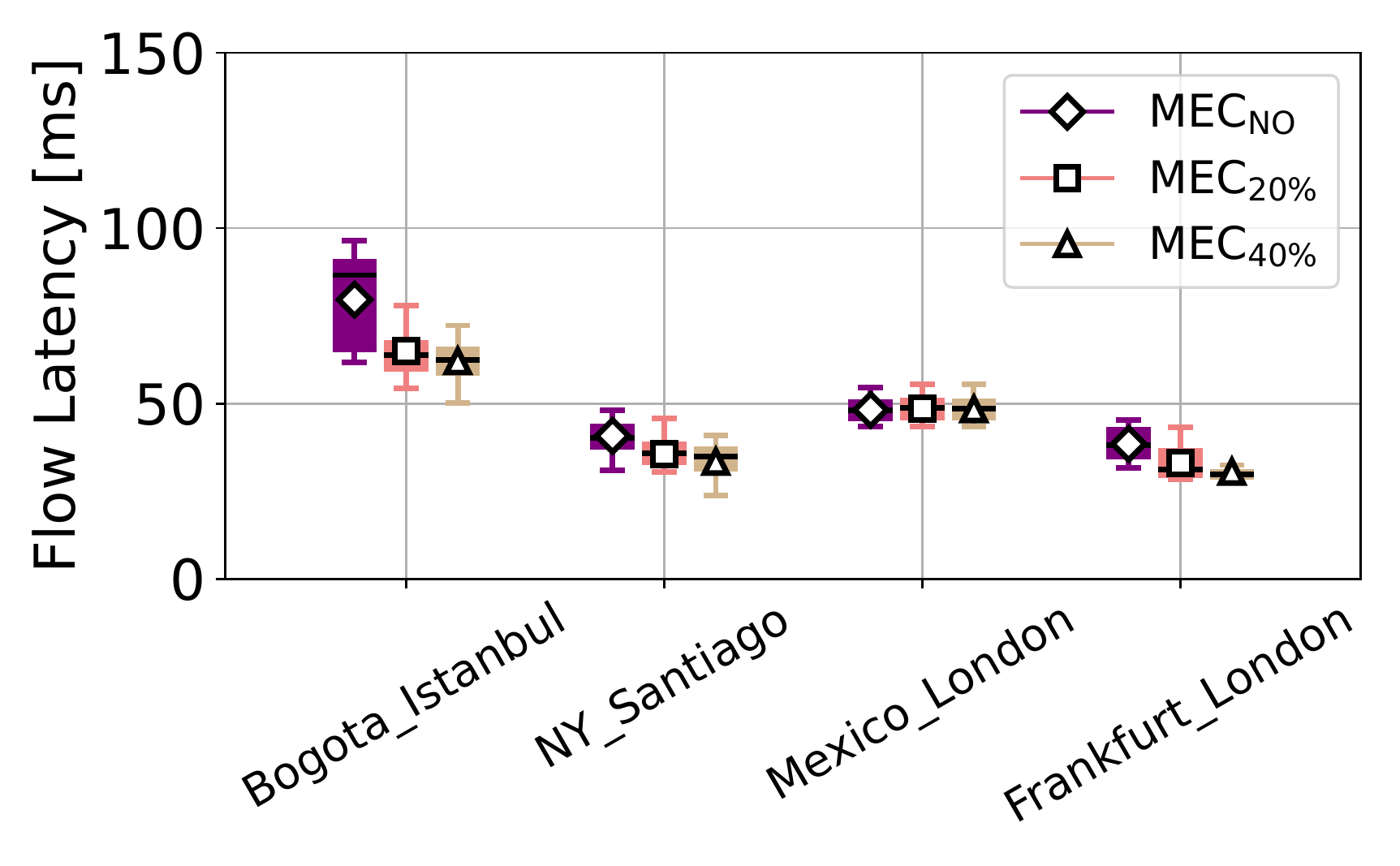}}\\
    \subfloat[Video Streaming]{%
    \label{fig:individual_aircrafts_Packet_latency_Video_Streaming_service}
    \includegraphics[width=0.45\textwidth]{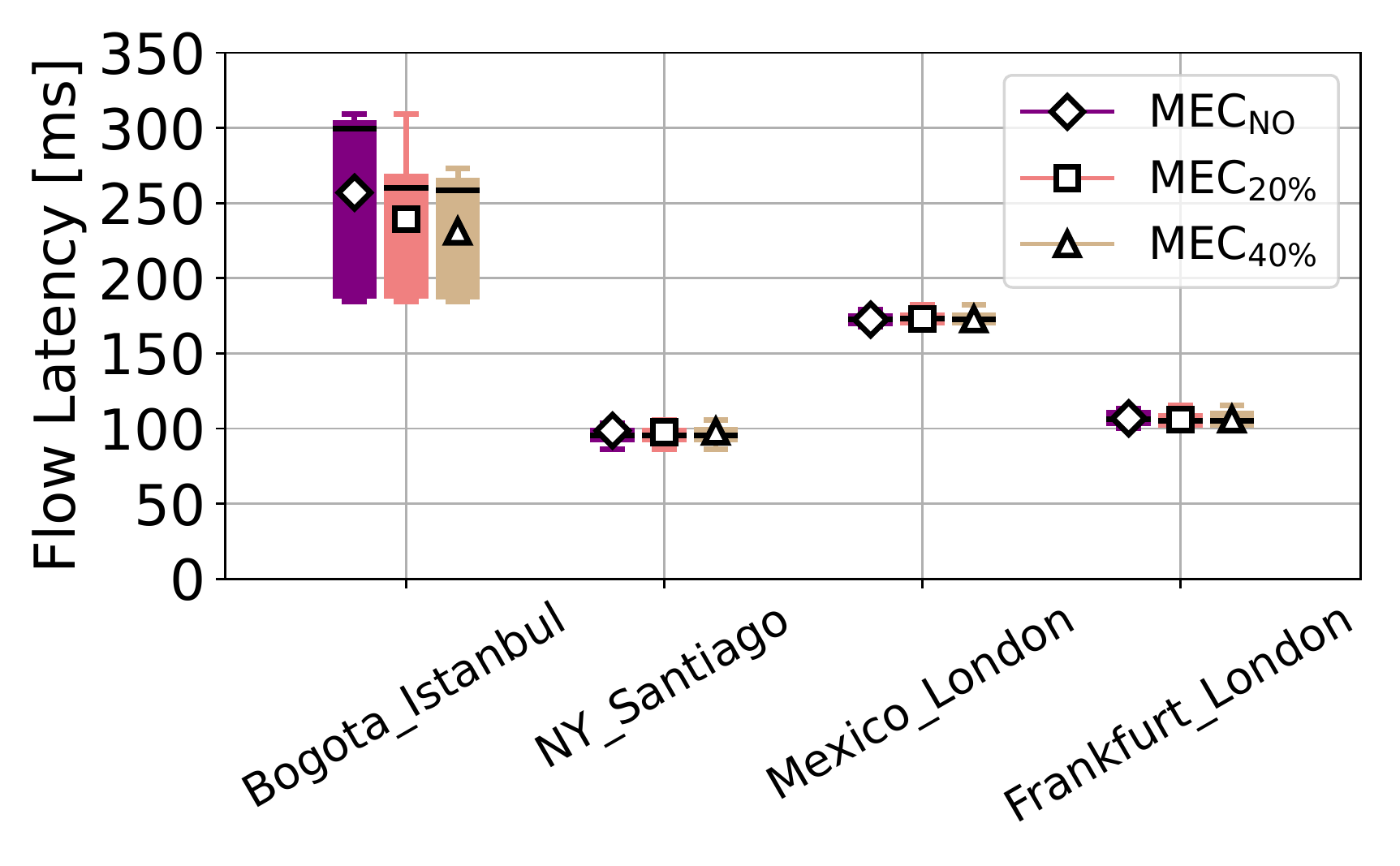}}
	\caption{Flow latency for different services for selected flights}
	\label{fig:selected_aircrafts}
\end{figure}

Fig.~\ref{fig:selected_aircrafts} shows the flow latency for selected flights and \ac{MEC} ratios over different applications. As the \ac{MEC} deployment ratio increases the flow latency decreases for all selected aircrafts. However, the performance gains differ highly between flights. This is due to various reasons like the changing number of passengers in aircrafts and the geographical location of the aircraft. Its location defines the gateway it will use. Due to uneven gateway distribution around the world, as shown in Table~\ref{tab:IridiumGateways}, some gateways need to cover a larger area and number of aircrafts, resulting in a higher latency for services to be routed to that gateway. Therefore, the network greatly benefits from deploying aerial \acp{MEC} around these gateways. This is the case for Bogota-Istanbul aircraft, which is reflected in the largest decrease in latency.

\begin{figure}[t]
	\centering
	\includegraphics[width=0.9\columnwidth]{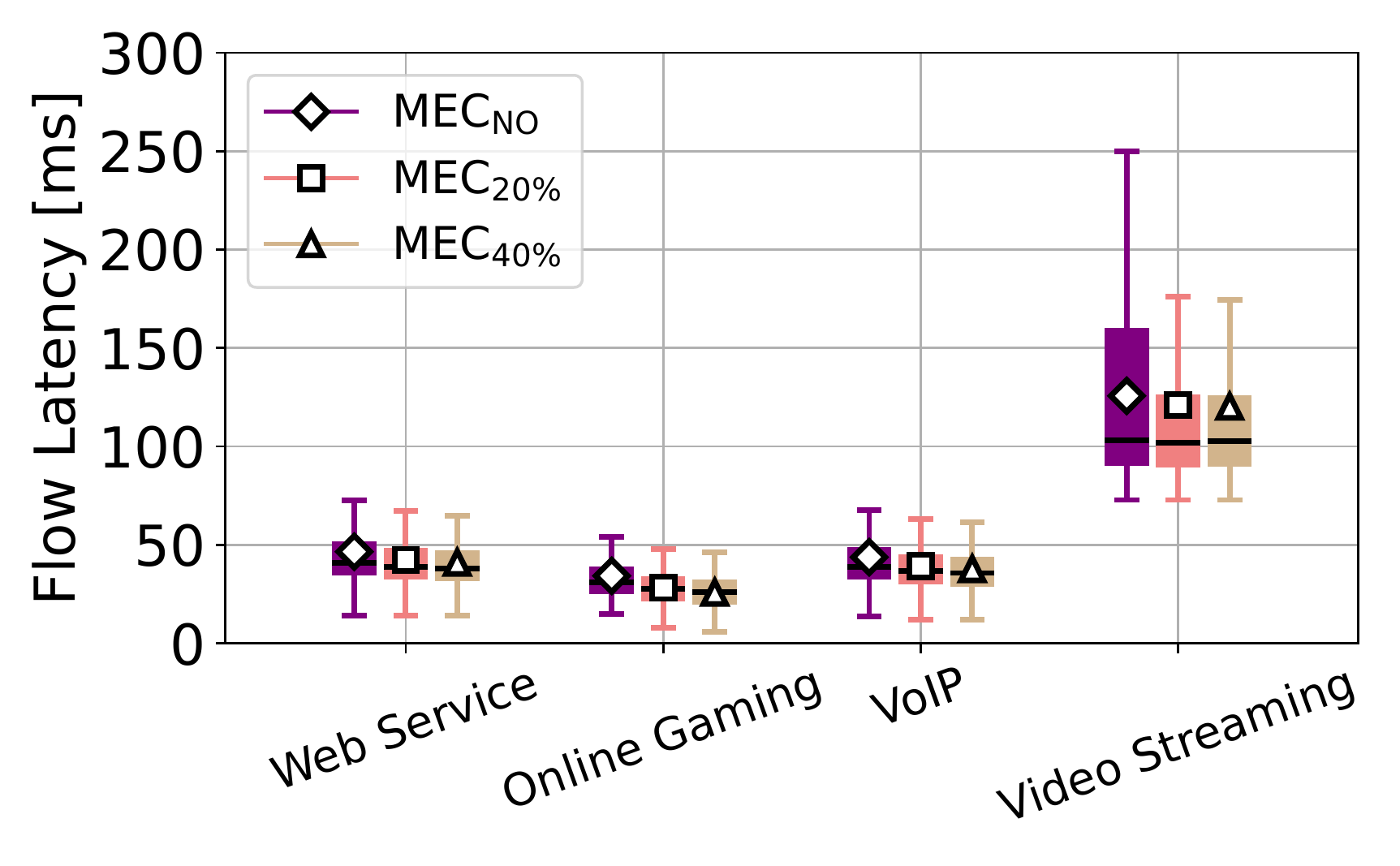}
	\caption{Overall system flow latency for aircrafts.}
	\label{fig:multiple_aircrafts_flow_latency}
\end{figure}

Fig.~\ref{fig:multiple_aircrafts_flow_latency}, shows the system flow latency over all aircrafts for different \ac{MEC} deployment ratios. 
Here, the spread of latencies decreases with increasing \ac{MEC} deployment ratio. The lower spread benefits \ac{QoS} guarantees for the different applications. On average, gaming benefits the most with 10.43\% and 16.01\% decrease in flow latency as the \ac{MEC} deployment ratio increases from 0 to 0.2 and from 0 to 0.4, respectively. In \ac{VoIP}, those decreases are 6.08\% and 8.32\%. In web services, the decreases are 5.38\% and 7.11\%. In video streaming, the decreases are 0.96\% and 0.41\%. These stated average percentage gains are heavily influenced by aircrafts that barely exhibit a latency reduction. The individual latency improvement for an aircraft can be considerably higher as Fig.~\ref{fig:selected_aircrafts} shows.
Since video streaming accounts for 67.5\% of the traffic in an aircraft, it has the highest flow size among the services and it becomes beneficial to optimize multiple smaller flows arising from the other services. Consequently, the decrease in flow latency is much smaller for video streaming. This is in accordance with our expectation from the services, since video streaming, in comparison to the other services, contains its own buffer and its \ac{QoS} is less sensitive to latency.

\subsection{Satellite Offloading} 
\label{results_satellite_offloading}
\begin{figure}[t]
\centering
    \subfloat[$\lambda=72$]{%
    \label{fig:individual_satellites_lambda_72_latency}
    \includegraphics[width=0.45\textwidth]{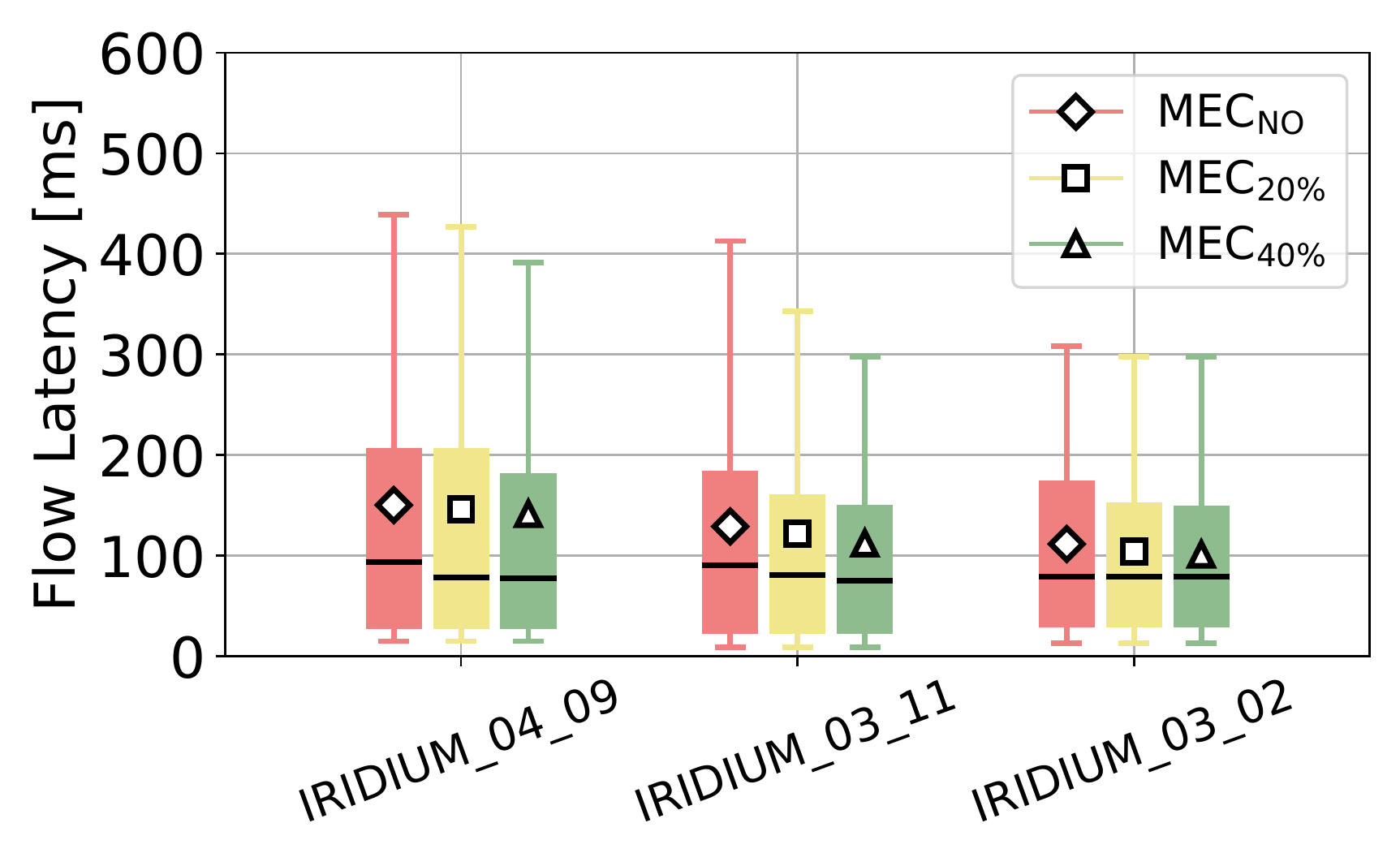}}\\
    \subfloat[$\lambda=76$]{%
    \label{fig:individual_satellites_lambda_76_latency}
    \includegraphics[width=0.45\textwidth]{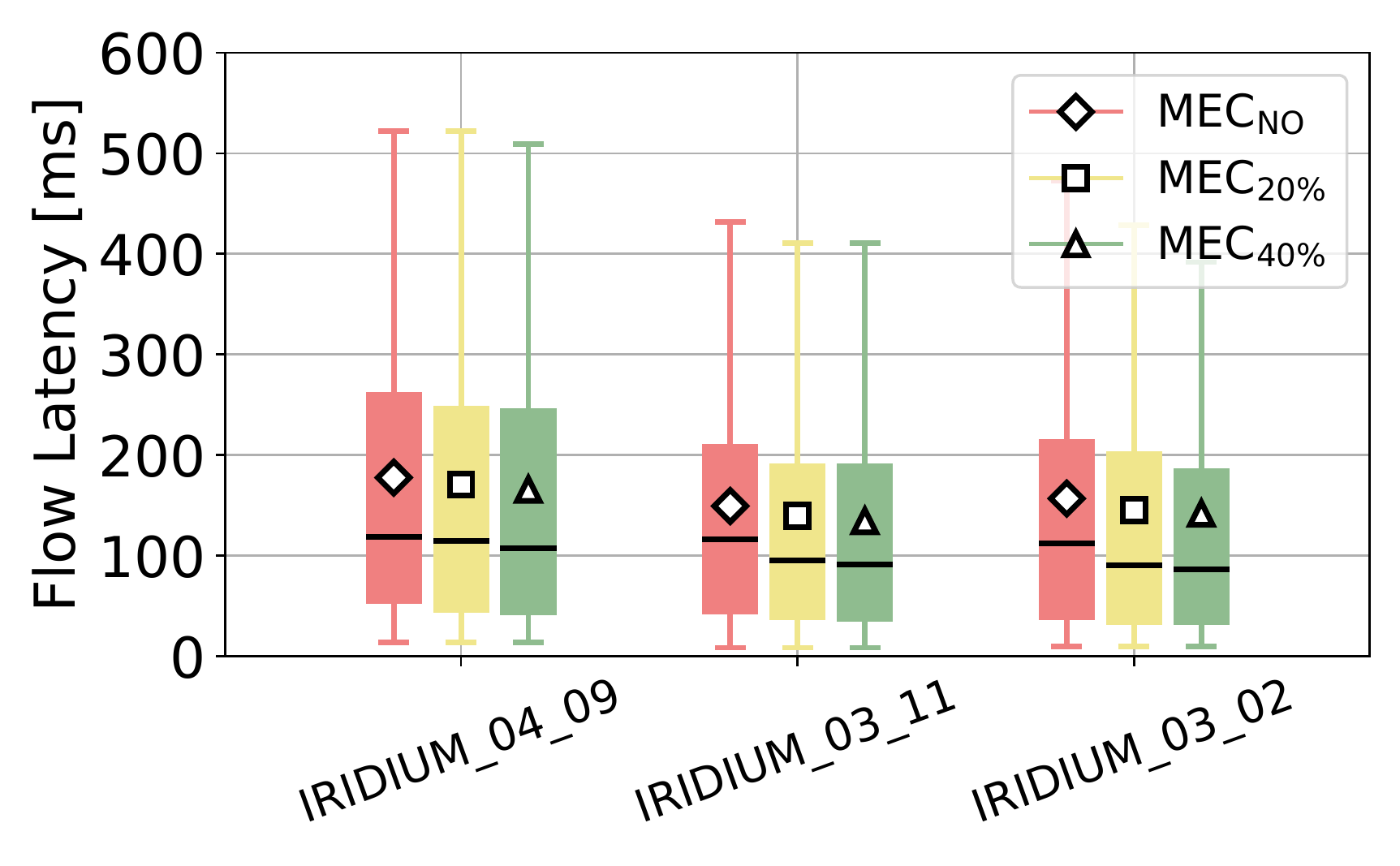}}\\
    \subfloat[$\lambda=80$]{%
    \label{fig:individual_satellites_lambda_80_latency}
    \includegraphics[width=0.45\textwidth]{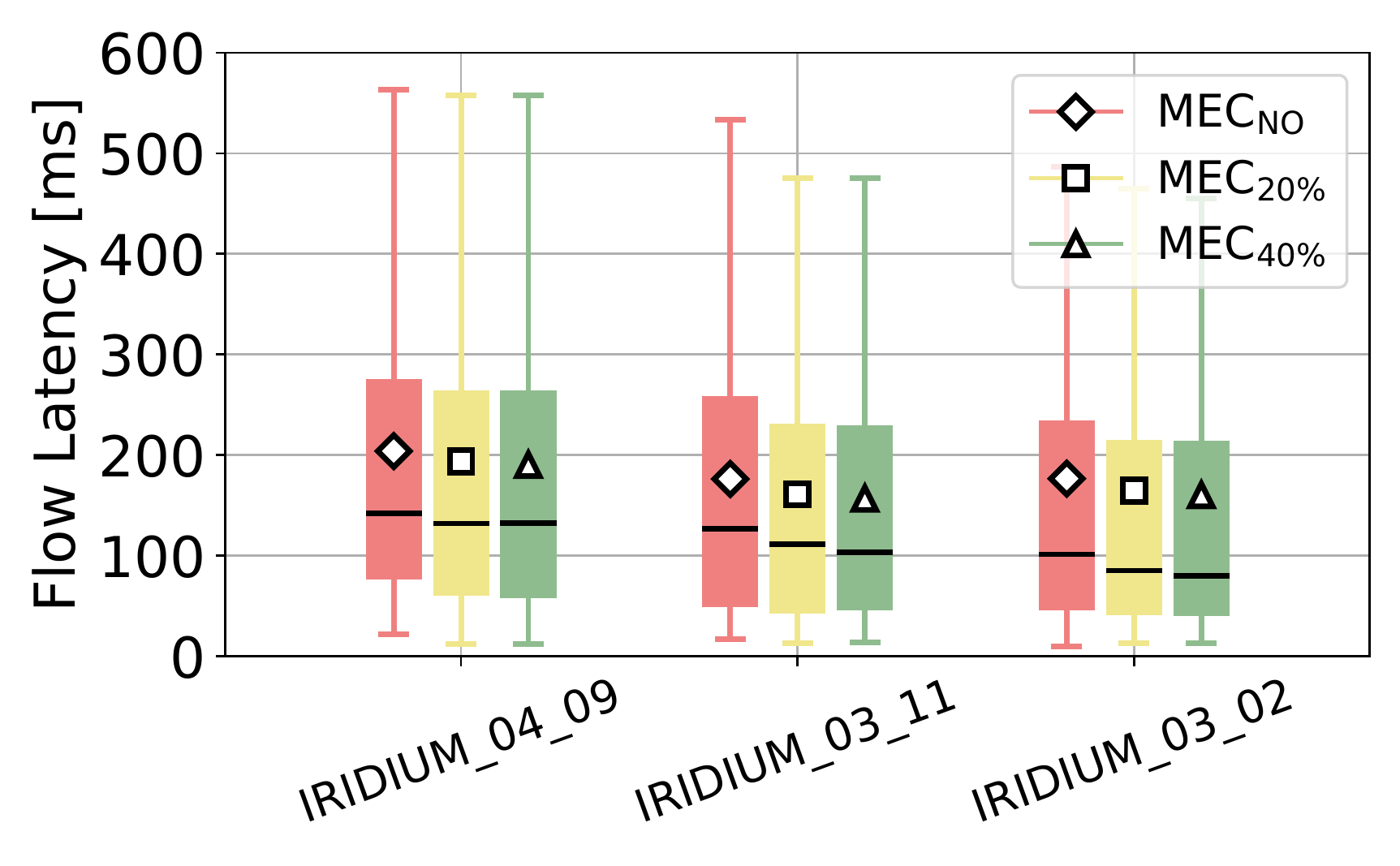}}\\
    \caption{Individual satellite flow latency for several arrivals task and \ac{MEC} probabilities}
	\label{fig:individual_satellites_latencies}
\end{figure}

\begin{figure}[t]
	\centering
	\includegraphics[width=0.9\columnwidth]{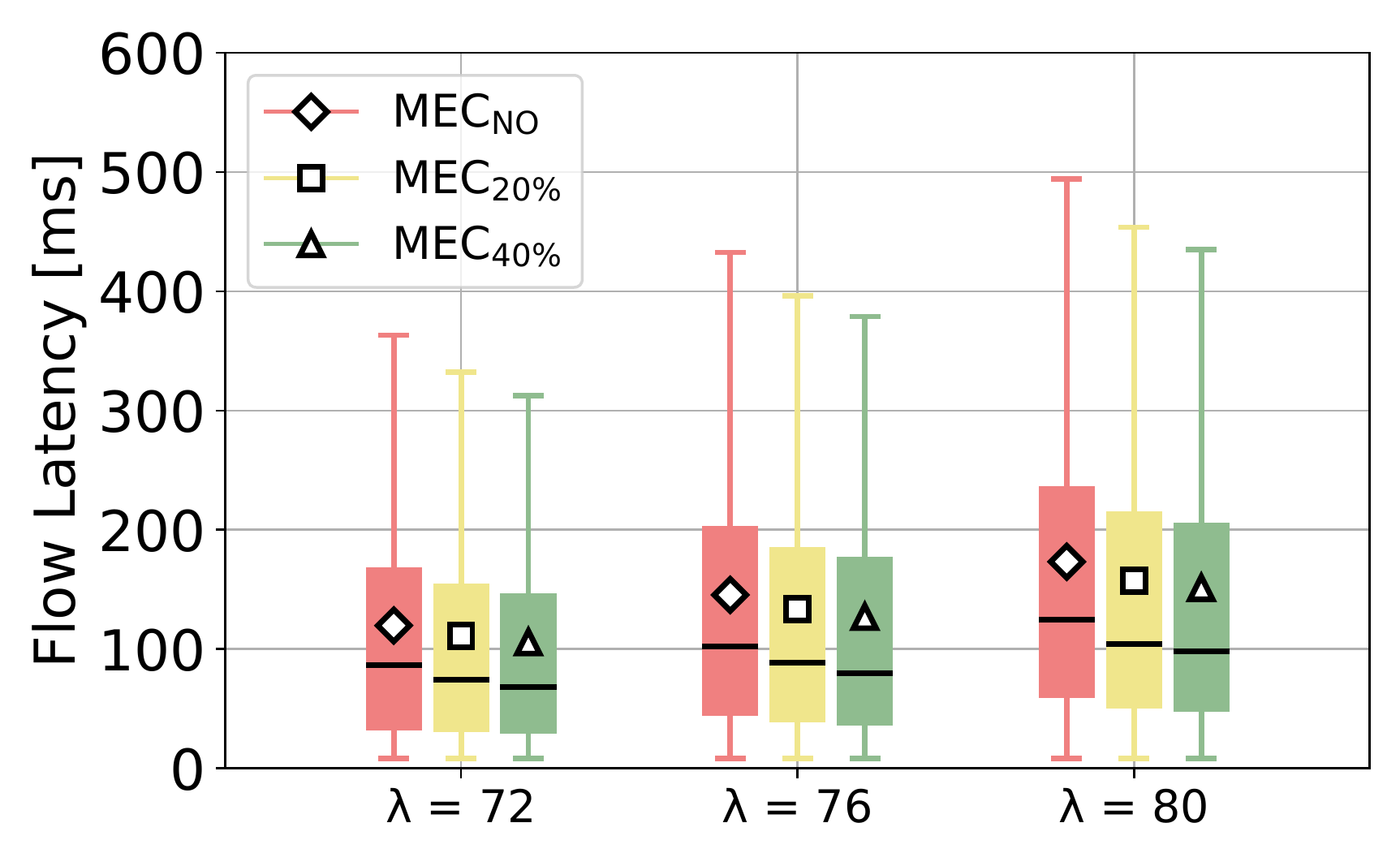}
	\caption{Overall system flow latency for satellites.}
	\label{fig:multiple_satellites_flow_latency}
\end{figure}

Looking further into the implications of the optimization in the use case of satellite task offloading, we demonstrate the benefits of the \ac{AA-MEC} for satellites.
Fig.~\ref{fig:individual_satellites_latencies} shows the flow latencies for three exemplary satellites for different aerial \ac{MEC} deployment ratios for varying task arrival rates $\lambda$. Similar to the airborne scenario, the average latency and the latency spread decrease with increasing \ac{MEC} deployment ratio. However, the performance gains vary less over the deployment ratios compared to the airborne scenario. This is due to the even distribution of satellites around the globe compared to aircrafts. 
Further, Fig.~\ref{fig:multiple_satellites_flow_latency} shows the overall system flow latency for varying task arrival rates $\lambda$. As expected, with increasing task arrival rate the flow latency increases. This can be counteracted by deploying more aerial \ac{MEC} nodes.
In case of $\lambda = 72$, the decreases in flow latencies are 14.47\% and 21.64\% as the \ac{MEC} deployment ratio goes from 0 to 0.2 and 0.4, respectively. For $\lambda = 76$ and $80$, those decreases are measured as 13.09\%, 22.09\% and 16.16\%, 21.38\%.
Analyzing these results in more detail, Table~\ref{tab:mec_utilization_satellite} confirms the airborne internet results, that with increasing \ac{MEC} deployment regions with a low count of satellite gateways profit the most. 
\begin{table}[t]
\caption{Flow distribution over different \ac{MEC} locations for varying aerial \ac{MEC} deployments for $\lambda = 80$}
\label{tab:mec_utilization_satellite}
\begin{center}
\setlength\tabcolsep{4pt}
\begin{tabular}{lccc}
\toprule
\textbf{\ac{MEC} Location} &\textbf{$\ac{MEC}_\text{NO}$} &\textbf{$\ac{MEC}_\text{20\%}$} &\textbf{$\ac{MEC}_\text{40\%}$}  \\
\midrule
Beijing & 15\% & 8\% & 6\%\\
Fairbanks & 6\% & 4\% & 4\%\\
Iqaluit & 8\% & 4\% & 4\%\\
Izhevsk & 8\% & 4\% & 4\%\\
Longyearbyen & 6\% & 4\% & 4\%\\
PuntaArenas & 27\% & 19\% & 16\%\\
Rome & 9\% & 6\% & 6\%\\
Tempe & 7\% & 4\% & 3\%\\
Wahiawa & 11\% & 9\% & 8\%\\
Yellowknife & 5\% & 3\% & 3\%\\
Aerial & 0\% & 34\% & 44\%\\
\bottomrule 
\end{tabular}
\end{center}
\end{table}

\begin{table}[t]
\caption{Bandwidth reduction compared to $\ac{MEC}_\text{NO}$}
\label{tab:mec_bandwidth}
\begin{center}
\setlength\tabcolsep{4pt}
\begin{tabular}{lccc}
\toprule
 & $\boldsymbol{\lambda=72}$ & $\boldsymbol{\lambda=76}$ & $\boldsymbol{\lambda=80}$ \\
\midrule
$\ac{MEC}_\text{20\%}$ & 11.93\% & 13.03\%	& 6.57\% \\
$\ac{MEC}_\text{40\%}$ & 17.63\% & 19.19\%	& 11.38\% \\
\bottomrule 
\end{tabular}
\end{center}
\end{table}
Optimizing for latency also influences the bandwidth requirement as shown in Table~\ref{tab:mec_bandwidth}. With increasing deployment of aerial \ac{MEC} nodes the utilized bandwidth of the network is reduced for any task arrival rate $\lambda$. However, the ratio between bandwidth reduction and \ac{MEC} deployment becomes nearly linear as soon as we saturate the local computing capabilities $\lambda=80$ and satellites are forced to offload. 

\section{Conclusion and Discussion}
\label{sec:conclusion}
In \aclp{MLN}, different layers support each other to provide better service in terms of coverage, latency and capacity. Moreover, \aclp{MLN} reduce the reliance on infrastructure by providing redundancy. Introducing \acl{MEC} on different layers allows for ubiquitous computing. Therefore, we propose the network architecture of \acl{AA-MEC} to boost network performance by bringing computing closer to the sources. This flexible architecture accounts for the dynamicity introduced by deploying aerial \ac{MEC}. 
In order to validate our approach, we implemented a three layer network consisting of a satellite layer with the Iridium-Next constellation, an aerial layer including various flights, and a terrestrial layer consisting of satellite gateways. In this network, the use case of airborne internet with different \acl{IFECS} and the use case of computational offloading from satellites are investigated. For these use cases we calculate the optimal \ac{MEC} destinations for task flows and the optimal route to these destinations.

In the use case of computational offloading for satellites, we can improve the flow latency by at least 56.03\% when comparing static networks with our dynamically optimized \ac{AA-MEC} network. Deploying aerial \acp{MEC} in addition to the terrestrial gateway \acp{MEC} reduces the flow latency by at least 13.09\%. Additionally, our \ac{AA-MEC} network brings forth at least 6.57\% decrease in occupied network bandwidth which results in less energy required for data transmission, increasing the network energy efficiency.
Further, the flow latency for \ac{IFECS} improve by 6.7\% from a static network to our \ac{AA-MEC}. The additional deployment of aerial \acp{MEC} on average results in a latency reduction of at least 5.71\%. Specifically, for gaming applications, the 10.43\% decrease in latency can be a deciding factor for the outcome of the game. Our proposed latency minimization problem formulation is especially suited for latency-critical applications with smaller capacity. This can easily be adapted for other configurations of services, for example a latency-critical service with a large bandwidth, by weighting each flow's latency in the objective function accordingly.

The improvements in latency demonstrated by our \ac{AA-MEC} can be further boosted by optimizing the selection of aircrafts that will deploy the \ac{MEC} server. Aircrafts requiring longer links to access the Iridium-Next gateways, like aircrafts flying over Africa, Australia or Oceans, greatly benefit from deploying \ac{MEC} servers on other aircrafts flying nearby. In scenarios where this is infeasible, considering the \aclp{HAP} in the aerial layer as \ac{MEC} nodes would prove beneficial. Furthermore, considering other satellite constellations would bring additional gateways into the picture, thereby influencing the distribution of \ac{MEC} nodes around the globe.

\begin{acronym}[Bash]
    \acro{5G}{Fifth Generation}
    \acro{6G}{Sixth Generation}
    \acro{AA-MEC}{Aerial-Aided Multi-Access Edge Computing}
    \acro{ABS}{Aerial Base Station}
    \acroplural{ABS}[ABSs]{Aerial Base Stations}
    \acro{AGC-MEC}{Air-Ground Collaborative Multi-Access Edge Computing}
    \acro{ARAN}{Aerial Radio Access Network}
    \acroplural{ARAN}[ARANs]{Aerial Radio Access Networks}
    \acro{DA2G}{Direct Air-to-Ground}
    \acro{ETSI}{European Telecommunications Standards Institute}
    \acro{HAP}{High-Altitude Platform}
    \acroplural{HAP}[HAPs]{High-Altitude Platforms}
    \acro{IFECS}{In-Flight Entertainment and Connectivity Service}
    \acroplural{IFECS}[IFECSs]{In-Flight Entertainment and Connectivity Services}
    \acro{IoT}{Internet of Things}
    \acroplural{ISL}[ISLs]{Inter-Satellite Links}
    \acro{LEO}{Low Earth Orbiting}
    \acro{LoS}{Line of Sight}
    \acro{MEC}{Multi-Access Edge Computing}
    \acro{MIPS}{Million Instructions per Second}
    \acro{MLN}{Multi-Layer Network}
    \acroplural{MLN}[MLNs]{Multi-Layer Networks}
    \acro{QoE}{Quality of Experience}
    \acro{QoS}{Quality of Service}
    \acro{RAN}{Radio Access Network}
    \acro{STK}{Systems Tool Kit}
    \acro{UAV}{Unmanned Aerial Vehicle}
    \acroplural{UAV}[UAVs]{Unmanned Aerial Vehicles}
    \acro{VoIP}{Voice over Internet Protocol}
\end{acronym}

\bibliographystyle{IEEEtaes}
\IEEEtriggeratref{27} 
\bibliography{references}

\begin{thebibliography}{10}
\providecommand{\url}[1]{#1}
\csname url@samestyle\endcsname
\renewcommand{\newblock}{\par}
\providecommand{\bibinfo}[2]{#2}
\providecommand{\BIBentrySTDinterwordspacing}{\spaceskip=0pt\relax}
\providecommand{\BIBentryALTinterwordstretchfactor}{4}
\providecommand{\BIBentryALTinterwordspacing}{\spaceskip=\fontdimen2\font plus
\BIBentryALTinterwordstretchfactor\fontdimen3\font minus
  \fontdimen4\font\relax}
\providecommand{\BIBforeignlanguage}[2]{{%
\expandafter\ifx\csname l@#1\endcsname\relax
\typeout{** WARNING: IEEEtran.bst: No hyphenation pattern has been}%
\typeout{** loaded for the language `#1'. Using the pattern for}%
\typeout{** the default language instead.}%
\else
\language=\csname l@#1\endcsname
\fi
#2}}
\providecommand{\BIBdecl}{\relax}
\BIBdecl

\bibitem{bernardos2021european}
C.~Bernardos and M.~Uusitalo
\newblock  {European vision for the 6G network ecosystem} \newblock
  \emph{Zenodo, Honolulu, HI, USA, Tech. Rep}, 2021.

\bibitem{dao2021survey}
N.-N. Dao, Q.-V. Pham, N.~H. Tu, T.~T. Thanh, V.~N.~Q. Bao, D.~S. Lakew, and
  S.~Cho
\newblock  {Survey on aerial radio access networks: Toward a comprehensive 6G
  access infrastructure} \newblock  \emph{IEEE Communications Surveys \&
  Tutorials}, vol.~23, no.~2, pp. 1193--1225, 2021.

\bibitem{jones2018recent}
H.~Jones
\newblock  {The recent large reduction in space launch cost} \newblock  In
  \emph{48th International Conference on Environmental Systems}, 2018.

\bibitem{chen2017caching}
M.~Chen, M.~Mozaffari, W.~Saad, C.~Yin, M.~Debbah, and C.~S. Hong
\newblock  Caching in the sky: Proactive deployment of cache-enabled unmanned
  aerial vehicles for optimized quality-of-experience \newblock  \emph{IEEE
  Journal on Selected Areas in Communications}, vol.~35, no.~5, pp. 1046--1061,
  2017.

\bibitem{pacheco2021towards}
L.~Pacheco, H.~Oliveira, D.~Ros{\'a}rio, Z.~Zhao, E.~Cerqueira, T.~Braun, and
  P.~Mendes
\newblock  Towards the future of edge computing in the sky: Outlook and future
  directions \newblock  In \emph{2021 17th International Conference on
  Distributed Computing in Sensor Systems (DCOSS)}. IEEE, 2021, pp. 220--227.

\bibitem{Medina2010}
D.~Medina, F.~Hoffmann, F.~Rossetto, and C.-H. Rokitansky
\newblock  {A crosslayer geographic routing algorithm for the airborne
  internet} \newblock  In \emph{2010 IEEE International Conference on
  Communications}. IEEE, 2010, pp. 1--6.

\bibitem{Varasteh2019Mobility}
A.~Varasteh, S.~Hofmann, N.~Deric, M.~He, D.~Schupke, W.~Kellerer, and C.~M.
  Machuca
\newblock  {Mobility-Aware Joint Service Placement and Routing in
  Space-Air-Ground Integrated Networks} \newblock  In \emph{ICC 2019 - 2019
  IEEE International Conference on Communications (ICC)}, 2019, pp. 1--7.

\bibitem{chen2021reinforcement}
Q.~Chen, W.~Meng, S.~Han, C.~Li, and H.-H. Chen
\newblock  Reinforcement learning-based energy-efficient data access for
  airborne users in civil aircrafts-enabled sagin \newblock  \emph{IEEE
  Transactions on Green Communications and Networking}, vol.~5, no.~2, pp.
  934--949, 2021.

\bibitem{Baek2019}
J.-y. Baek, G.~Kaddoum, S.~Garg, K.~Kaur, and V.~Gravel
\newblock  {Managing Fog Networks using Reinforcement Learning Based Load
  Balancing Algorithm} \newblock  In \emph{2019 IEEE Wireless Communications
  and Networking Conference (WCNC)}, 2019, pp. 1--7.

\bibitem{Wang2018}
Y.~Wang, J.~Zhang, X.~Zhang, P.~Wang, and L.~Liu
\newblock  {A Computation Offloading Strategy in Satellite Terrestrial Networks
  with Double Edge Computing} \newblock  In \emph{2018 IEEE International
  Conference on Communication Systems (ICCS)}, 2018, pp. 450--455.

\bibitem{Cheng2019}
N.~Cheng, F.~Lyu, W.~Quan, C.~Zhou, H.~He, W.~Shi, and X.~Shen
\newblock  {Space/aerial-assisted computing offloading for IoT applications: A
  learning-based approach} \newblock  \emph{IEEE Journal on Selected Areas in
  Communications}, vol.~37, no.~5, pp. 1117--1129, 2019.

\bibitem{Zhang2019}
Z.~Zhang, W.~Zhang, and F.-H. Tseng
\newblock  {Satellite Mobile Edge Computing: Improving QoS of High-Speed
  Satellite-Terrestrial Networks Using Edge Computing Techniques} \newblock
  \emph{IEEE Network}, vol.~33, no.~1, pp. 70--76, 2019.

\bibitem{yu2021ec}
S.~Yu, X.~Gong, Q.~Shi, X.~Wang, and X.~Chen
\newblock  Ec-sagins: Edge-computing-enhanced space--air--ground-integrated
  networks for internet of vehicles \newblock  \emph{IEEE Internet of Things
  Journal}, vol.~9, no.~8, pp. 5742--5754, 2021.

\bibitem{inmarsat_webpage}
\BIBentryALTinterwordspacing

\newblock \emph{{Inmarsat Orchestra}}. Accessed: 2021-10-15. [Online].
  Available: \url{https://www.inmarsat.com/en/about/technology/orchestra.html}
\BIBentrySTDinterwordspacing

\bibitem{Chen2021}
Q.~Chen, L.~Yang, X.~Liu, J.~Guo, S.~Wu, and X.~Chen
\newblock  {Multiple gateway placement in large-scale constellation networks
  with inter-satellite links} \newblock  \emph{International Journal of
  Satellite Communications and Networking}, vol.~39, no.~1, pp. 47--64, 2021.

\bibitem{Zhen2021}
Z.~Qin, H.~Wang, Y.~Qu, H.~Dai, and Z.~Wei
\newblock  {Air-Ground Collaborative Mobile Edge Computing: Architecture,
  Challenges, and Opportunities} \newblock  \emph{arXiv preprint
  arXiv:2101.07930}, 2021.

\bibitem{Osoro2021}
O.~B. Osoro and E.~J. Oughton
\newblock  {A Techno-Economic Framework for Satellite Networks Applied to Low
  Earth Orbit Constellations: Assessing Starlink, OneWeb and Kuiper} \newblock
  \emph{IEEE Access}, vol.~9, pp. 141\,611--141\,625, 2021.

\bibitem{Liu2018}
J.~Liu, Y.~Shi, Z.~M. Fadlullah, and N.~Kato
\newblock  {Space-air-ground integrated network: A survey} \newblock
  \emph{IEEE Communications Surveys \& Tutorials}, vol.~20, no.~4, pp.
  2714--2741, 2018.

\bibitem{Saeed2021}
N.~Saeed, H.~Almorad, H.~Dahrouj, T.~Y. Al-Naffouri, J.~S. Shamma, and M.-S.
  Alouini
\newblock  {Point-to-point communication in integrated satellite-aerial 6G
  networks: State-of-the-art and future challenges} \newblock  \emph{IEEE Open
  Journal of the Communications Society}, 2021.

\bibitem{IridiumConstellation}
\BIBentryALTinterwordspacing

\newblock \emph{{Iridium Network}}. Accessed: 2022-04-14. [Online]. Available:
  \url{https://www.iridium.com/}
\BIBentrySTDinterwordspacing

\bibitem{Del2018}
I.~del Portillo, B.~Cameron, and E.~Crawley
\newblock  {Ground segment architectures for large LEO constellations with
  feeder links in EHF-bands} \newblock  In \emph{2018 IEEE Aerospace
  Conference}. IEEE, 2018, pp. 1--14.

\bibitem{iridium_gateways}
I.~C. Inc.
\newblock  2016 annual report \newblock  IRIDIUM Communications Inc., Tech.
  Rep., 2016.

\bibitem{etsi_arch}
\emph{{Mobile Edge Computing (MEC) Framework and Reference Architecture}}
\newblock ETSI Group Specification MEC 003, Rev. 1.1.1, 03 2016.

\bibitem{ETSI_MEC}
\BIBentryALTinterwordspacing

\newblock \emph{{ETSI - Multi-access Edge Computing (MEC)}}. Accessed:
  2022-01-16. [Online]. Available:
  \url{https://www.etsi.org/technologies/multi-access-edge-computing}
\BIBentrySTDinterwordspacing

\bibitem{Wang2021ntn6G}
D.~Wang, M.~Giordani, M.-S. Alouini, and M.~Zorzi
\newblock  {The Potential of Multilayered Hierarchical Nonterrestrial Networks
  for 6G: A Comparative Analysis Among Networking Architectures} \newblock
  \emph{IEEE Vehicular Technology Magazine}, vol.~16, no.~3, pp. 99--107, 2021.

\bibitem{Giordani2021}
M.~Giordani and M.~Zorzi
\newblock  {Non-Terrestrial Networks in the 6G Era: Challenges and
  Opportunities} \newblock  \emph{IEEE Network}, vol.~35, no.~2, pp. 244--251,
  2021.

\bibitem{etsi_spesification}
\emph{{ETSI TS 120 203}}
\newblock ETSI Technical Specification 123 203, Rev. 13.6.0, 03 2016.

\bibitem{savi2015impactservices}
M.~Savi, M.~Tornatore, and G.~Verticale
\newblock  Impact of processing costs on service chain placement in network
  functions virtualization \newblock  In \emph{2015 IEEE conference on network
  function virtualization and software defined network (NFV-SDN)}. IEEE, 2015,
  pp. 191--197.

\bibitem{Garcia2018}
J.~Garcia and A.~Brunstrom
\newblock  {Clustering-based separation of media transfers in DPI-classified
  cellular video and VoIP traffic} \newblock  In \emph{2018 IEEE Wireless
  Communications and Networking Conference (WCNC)}, 2018, pp. 1--6.

\bibitem{Jorgensen2020}
P.~J. Jorgensen, O.~F. Awad, and R.~H. Bishop
\newblock  {Analysis and Design of a Sub-Optimal MEKF for Low Earth Orbit
  Attitude Estimation Using a Radically Inexpensive MEMS IMU} \newblock  In
  \emph{Small Satellite Conference}, 2020.

\bibitem{agi_stk}
\BIBentryALTinterwordspacing

\newblock \emph{{AGI. Systems tool kit (stk)}}. Accessed: 2021-15-10. [Online].
  Available: \url{https://www.agi.com/products/engineering-tools}
\BIBentrySTDinterwordspacing

\bibitem{papa2020design}
A.~Papa, T.~De~Cola, P.~Vizarreta, M.~He, C.~Mas-Machuca, and W.~Kellerer
\newblock  {Design and evaluation of reconfigurable SDN LEO constellations}
  \newblock  \emph{IEEE Transactions on Network and Service Management},
  vol.~17, no.~3, pp. 1432--1445, 2020.

\bibitem{Sthapit2021}
S.~Sthapit, S.~Lakshminarayana, L.~He, G.~Epiphaniou, and C.~Maple
\newblock  {Reinforcement Learning for Security Aware Computation Offloading in
  Satellite Networks} \newblock  \emph{IEEE Internet of Things Journal}, pp.
  1--1, 2021.

\bibitem{Rodrigues2017}
T.~G. Rodrigues, K.~Suto, H.~Nishiyama, and N.~Kato
\newblock  {Hybrid Method for Minimizing Service Delay in Edge Cloud Computing
  Through VM Migration and Transmission Power Control} \newblock  \emph{IEEE
  Transactions on Computers}, vol.~66, no.~5, pp. 810--819, 2017.

\bibitem{gurobi}
\BIBentryALTinterwordspacing
{Gurobi Optimization, LLC}
\newblock  {Gurobi Optimizer Reference Manual} \newblock  2022. [Online].
  Available: \url{https://www.gurobi.com}
\BIBentrySTDinterwordspacing

\end{thebibliography}
\end{document}